\documentclass[aps,twocolumn,preprintnumbers,nofootinbib,prd,superscriptaddress,10pt]{revtex4-1}

\usepackage[utf8]{inputenc}

\usepackage{graphicx,amssymb,amsmath,amsthm,amsfonts,epsfig,epsf}
\usepackage[usenames,dvipsnames,svgnames,table]{xcolor}
\usepackage{bm}
\usepackage{dcolumn}
\usepackage{latexsym}
\usepackage{rotating}
\usepackage{enumerate}
\usepackage{tensor,multirow}
\usepackage{url}
\usepackage[linktocpage]{hyperref}
\usepackage{braket}
\usepackage{diagbox}

\def\be{\begin{equation}}
\def\ee{\end{equation}}
\newcommand{\beq}{\begin{eqnarray}}
\newcommand{\eeq}{\end{eqnarray}}

\def\ba{\begin{align}}
\def\ea{\end{align}}

\begin{document}
\title{Black hole binaries and light fields: Gravitational molecules}

\author{Taishi Ikeda}
\affiliation{Dipartimento di Fisica, 
``Sapienza'' Universit\'{a}
di Roma, Piazzale Aldo Moro 5, 00185, Roma, Italy}
\affiliation{CENTRA, Departamento de F\'{\i}sica, Instituto Superior T\'ecnico -- IST, Universidade de Lisboa -- UL,
Avenida Rovisco Pais 1, 1049 Lisboa, Portugal}

\author{Laura Bernard}
\affiliation{LUTH, Observatoire de Paris, PSL Research University, CNRS, Universit\'e de Paris, 5 place Jules Janssen, 92195 Meudon, France}

\author{Vitor Cardoso}
\affiliation{CENTRA, Departamento de F\'{\i}sica, Instituto Superior T\'ecnico -- IST, Universidade de Lisboa -- UL,
Avenida Rovisco Pais 1, 1049 Lisboa, Portugal}

\author{Miguel Zilh\~ao}
\affiliation{CENTRA, Departamento de F\'{\i}sica, Instituto Superior T\'ecnico -- IST, Universidade de Lisboa -- UL,
Avenida Rovisco Pais 1, 1049 Lisboa, Portugal}

\begin{abstract}
We show that light scalars can form quasibound states around binaries. In the nonrelativistic regime, these states are formally described by the quantum-mechanical Schr\"odinger equation
for a one-electron heteronuclear diatomic molecule.
We performed extensive numerical simulations of scalar fields around black hole binaries
showing that a scalar structure condenses around the binary -- we dub these states ``gravitational molecules''.
We further show that these are well described by the perturbative, nonrelativistic description.
\end{abstract}

\maketitle

% \tableofcontents

%%%%%%%%%%%%%%%%%%%%%%%%%%%%%%%%%%%%%%%%%%%%%%%%%%%%%%%%%%%%%%%%%%%%%%%%%%%%%
\section{Introduction}
%%%%%%%%%%%%%%%%%%%%%%%%%%%%%%%%%%%%%%%%%%%%%%%%%%%%%%%%%%%%%%%%%%%%%%%%%%%%%

More than one century after Einstein wrote down the field equations of general
relativity, black holes (BHs) remain one of its most outstanding and intriguing
predictions. Among all of its features, the inherent simplicity of stationary BHs is
possibly the most remarkable one: just two numbers (mass and angular momentum)
suffice to fully characterize these objects in vacuum~\cite{Bekenstein:1971hc,Robinson:1975bv,Bekenstein:1996pn,Chrusciel:2012jk}.

This simplicity and fundamental nature has led to analogies being drawn between
BHs in general relativity and the hydrogen atom in quantum mechanics. 
In fact, in the context of fundamental fields in curved spacetime, BHs do behave
as atoms: massive scalar fields can form long-lived states which are, in a certain limit, mathematically described
by the nonrelativistic Schr\"odinger equation for the hydrogen atom~\cite{Brito:2015oca,Detweiler:1980uk,Arvanitaki:2010sy,Baumann:2019eav};
such states have been dubbed ``gravitational atoms''~\cite{Baumann:2019eav}.

The horizon of non-spinning BHs acts as a dissipative surface, and hence these scalar states are in general ``quasibound''.
When the host BH is spinning these configurations may grow via superradiance, extracting a substantial fraction of the BH rotation energy to a bosonic
``cloud'' in the BH exterior. The process slows the BH spin down and releases monochromatic gravitational waves, giving rise to very particular imprints.
These states may form, through a different process, as a consequence of boson star collisions~\cite{Sanchis-Gual:2020mzb} or collisions between axion stars and BHs~\cite{Clough:2018exo}.
Thus BHs can be used as efficient particle detectors of ultralight fields across a wide range of mass~\cite{Brito:2015oca,Brito:2014wla,Yoshino:2013ofa}.
For fine-tuned conditions, the states can become truly bound states, and new BH solutions become possible~\cite{Herdeiro:2014goa}.

Black hole binaries were recently shown to have characteristic vibration
modes~\cite{Bernard:2019nkv}. Together with the above discussion on gravitational atoms, one is led to ask whether
the program can be taken a step further: Does it make sense to talk about ``gravitational molecules''?
Recent work using effective field theory techniques indicates that quasibound states of light scalars engulfing binaries could exist~\cite{Wong:2020qom}. We explore this question further here both analytically and numerically, and give a positive answer providing (under certain mild conditions), an equivalence between BH binaries and dihydrogen molecules.

We will study this issue by looking at the dynamics of massive scalar fields
in a BH binary (BHB) spacetime. We thus consider a Klein-Gordon scalar, governed by the equation
\begin{equation}
\square \phi = \mu^2 \phi\,,\label{eq:KG}
\end{equation}
in a nontrivial background describing a binary---in particular, a BHB. We always neglect backreaction of the field in the spacetime geometry, which in all but extreme situations  should be a very accurate approximation. Here, the mass parameter $\mu$ is related to the boson mass $m_B=\hbar \mu$.

%
%This work is structured as follows. In Sec.~\ref{section:non_relativistic} we
%derive the nonrelativistic approximation of the system under consideration,
%establish its formal equivalence with that of the dihydrogen molecule in
%Quantum Mechanics and treat the corresponding equations perturbatively. In
%Sec.~\ref{sec:numerics} we present the setup for our (non-perturbative)
%numerical treatment, and show the corresponding results in
%Sec.~\ref{sec:results}. We conclude in Sec.~\ref{sec:conclusions}.

We use geometric units $G=c=1$ and a $(-+++)$ convention for the metric
signature throughout.

%%%%%%%%%%%%%%%%%%%%%%%%%%%%%%%%%%%%%%%%%%%%%%%%%%%%
\section{Nonrelativistic scalars around binaries}
\label{section:non_relativistic}
%%%%%%%%%%%%%%%%%%%%%%%%%%%%%%%%%%%%%%%%%%%%%%%%%%%%
We first explore the problem in a nonrelativistic setting. This means that the spacetime is taken to be weakly curved,
and could describe, for instance, two noncompact stars. It also means that the rest mass of the scalar dominates over its kinetic energy.

%%%%%%%%%%%%%%%%%%%%%%%%%%%%%%%%%%%%%%%%%%%%%%%%%%% 
\subsection{Equivalence with dihydrogen molecules}
\label{sec:equivalence_molecule}
%%%%%%%%%%%%%%%%%%%%%%%%%%%%%%%%%%%%%%%%%%%%%%%%%%%
In a nonrelativistic setting, the Klein-Gordon equation around a single BH reduces to the Schr\"odinger equation~\cite{Brito:2015oca,Arvanitaki:2010sy,Baumann:2019eav}. 
There is thus a quasibound state structure for scalar fields around a BH which is identical to the spectrum of the hydrogen atom.
Let us now consider the nonrelativistic limit of a scalar field around a BHB.
To lowest order in a post-Newtonian expansion, the geometry of a binary (including that of a BHB, if we are not interested in near-horizon phenomena) can be written
in the form
\begin{equation}
ds^{2}=-\left(1+2\Phi_{N}\right)dt^{2}+(1-2\Phi_{N})\delta_{ij}dx^{i}dx^{j}\,,  \label{eq:metric}
\end{equation}
where
\begin{equation}
\Phi_{N}(t,x^{i})=-\frac{M_{1}}{|\vec{r}-\vec{r}_{1}(t)|}-\frac{M_{2}}{|\vec{r}-\vec{r}_{2}(t)|}\,,\label{eq:newton}
\end{equation}
is Newton's potential. Here, $M_{i} (i=1,2)$ are the individual component masses, and $\vec{r}_{i}(t)$ are their position vector.
There are higher-order terms which depend on the specifics of the system, and which become relevant for relativistic
and strongly gravitating systems, but which do not affect the physics we are interested in.

Using standard nonrelativistic limit procedures, we define the complex field $\Psi(t,\vec{r})$ as
\begin{equation}
\phi=\frac{1}{\sqrt{2\mu}}\left(\Psi e^{-i\mu t}+\Psi^{\ast} e^{i\mu t}\right)\,.
\end{equation}
Note that the Klein-Gordon field $\phi$ is assumed to be real throughout this work. The field $\Psi$ is, in general, complex.
% In the nonrelativistic limit, for very small scalar mass $\mu$ and extended scalar configurations, the field $\Psi$ is time-independent. We now wish to understand corrections to this regime.
Assuming that the binary components are widely separated and that the angular
frequency is so small that time-dependent terms in the Newtonian potential can
be neglected, the Klein-Gordon equation~(\ref{eq:KG}) reduces to the Schr\"odinger equation
\begin{equation}
  \label{eq:schrod}
  i\partial_t\Psi(t,x^i)= \left(-\frac{\nabla^{2}}{2\mu} + \mu \, \Phi_{N} \right) \Psi(t,x^i) \,,
\end{equation}
where we neglect the subleading (for weakly gravitating, nonrelativistic systems) terms 
\[
% \frac{2}{\mu}\dot{\Phi}_{N}(\dot{\Psi}-i\mu\Psi)+\frac{1}{2\mu}\ddot{\Psi}-\frac{2}{\mu}\Phi_{N}\nabla^{2}\Psi \,.
  \frac{2}{\mu}\partial_t{\Phi}_{N}\partial_t{\Psi}, \quad
  2 i \partial_t{\Phi}_{N}\Psi, \quad
  \frac{\partial^2_t{\Psi}}{2\mu}, \quad
  \frac{2}{\mu}\Phi_{N}\nabla^{2}\Psi \,,
\]
We can also recover Eq.~\eqref{eq:schrod} via a Lagrangian approach~\cite{Baumann:2018vus}.

Equation~(\ref{eq:schrod}) is written in the lab frame,
$x^{\mu} = (t, r, \theta, \varphi) $. It will be useful to write it in the binary rest
frame (corotating frame), $\bar x^\mu = (\bar t, \bar r, \bar \theta, \bar \varphi)$,
which we can do with the usual coordinate transformation
\begin{equation}
\label{eq:torestframe}
\partial_{t} = \partial_{\bar t} - \Omega \partial_{\bar \varphi} \,, \qquad
\partial_{\varphi}= \partial_{\bar \varphi} \,,
\end{equation}
where $\partial_{\bar \varphi} = - \bar y \partial_{\bar x} +  \bar x \partial_{\bar y}$ spans the orbital plane, and $\Omega$ is the BHB orbital angular velocity.
The frames are related through
\begin{equation}
\bar t  =  t \,, \qquad
\bar r =  r  \,, \qquad
\bar \theta = \theta \,, \qquad
\bar \varphi = \varphi - \Omega \, t \,.
\label{eq:xtoxbar}
\end{equation}
Denoting $\bar \Psi(\bar t, \bar r, \bar \theta, \bar \varphi) \equiv \Psi(\bar t, \bar r, \bar \theta, \bar \varphi + \Omega \bar t) = \Psi(t, r, \theta, \varphi)$, and
remembering that $\nabla^{2} = \bar{\nabla}^{2}$,
we can then rewrite Eq.~\eqref{eq:schrod} in the corotating frame as
\begin{equation}
i \partial_{\bar t} \bar \Psi(\bar t,\bar{x}^i) = H_0 \bar \Psi(\bar t,\bar{x}^i)
+ i \Omega \partial_{\bar \varphi} \bar \Psi(\bar t,\bar{x}^i) \,,\label{eq:schrod-pert}
\end{equation}
where
\begin{equation}
H_0 = -\frac{1}{2\mu} \bar \nabla^{2} + V\,,\label{eq:ham}
\end{equation}
and the (time-independent) potential $V$ is given by
\begin{equation}
V =-\frac{\mu M_1}{r_1}-\frac{\mu M_2}{r_2} \,,\label{eq:potential}
\end{equation}
where $r_{1,2}=\sqrt{(\bar x\mp a)^2 + \bar y^2 + \bar z^2}$ is the distance to BH $i$, and $\bar{x} = \pm a$ are the positions of each BH.

Equation~\eqref{eq:schrod-pert} can be treated perturbatively when $\Omega$ is
small. Let us first consider the unperturbed system,
\[
i \partial_{\bar t} \bar \Psi = H_0 \bar \Psi \,.
\]
Since the potential $V$ is not time dependent, we can consider the energy
eigenstate problem; writing
\be
\bar \Psi(\bar t,\bar{x}^i)=\bar \psi(\bar{x}^i)e^{-i\bar{E} \bar t}\,,
\ee
we then have
\begin{equation}
\bar{E} \bar \psi=-\frac{1}{2\mu}\bar{\nabla}^{2}\bar \psi + V \bar \psi\,.\label{eq:eigenvalue}
\end{equation}
Introducing prolate spheroidal coordinates
\begin{equation}
\begin{aligned}
\bar x(\xi,\eta,\chi )&= a \, \eta \, \xi \\
\bar y(\xi,\eta,\chi )&= a \sqrt{1-\eta ^2} \sqrt{\xi ^2-1} \sin (\chi )\\
\bar z(\xi,\eta,\chi )&= a \sqrt{1-\eta ^2} \sqrt{\xi ^2-1} \cos (\chi )
\end{aligned} \,,
\label{eq:prolate-def}
\end{equation}
where $2a$ is the separation between the two BHs on the $\bar x$ axis and
$-1\leq \eta\leq1$, $1\leq\xi$, and $0\leq \chi<2\pi$,
Eq.~(\ref{eq:eigenvalue}) becomes separable. Using the ansatz
\begin{equation}
\bar \psi(\xi,\eta,\chi)=\frac{e^{im_{\chi}\chi}}{\sqrt{2\pi}}R(\xi)S(\eta)\,,\label{eq:eingenstate}
\end{equation}
with $m_{\chi}=0,\pm 1,\pm 2,\ldots$, we then find
%
% \begin{widetext}
\begin{subequations}
\begin{align}
&0=\partial_{\eta}\left((1-\eta^{2})\partial_{\eta}S\right)\nonumber\\
& \quad+\left(A-2\mu a^{2}\bar{E}\eta^{2}+2a\mu\Delta\alpha\eta+\frac{m_{\chi}^2}{\eta^{2}-1}\right)S\,,\\
&0=\partial_{\xi}\left((\xi^{2}-1)\partial_{\xi}R\right)\nonumber\\
&\quad+\left(-A-\frac{m_{\chi}^2}{\xi^{2}-1}+2\alpha \mu a\xi+2\mu a^{2}\bar{E}\xi^{2}\right)R\,,
\end{align}
\label{eq:spheroidals}
\end{subequations}
where we define
\[
  \alpha_i=M_i\mu\,, \qquad
  \alpha=\alpha_{1}+\alpha_{2}\,, \qquad
  \Delta\alpha=\alpha_{1}-\alpha_{2} \,.
\]
Here, $A$ is a separation constant. $\bar{E}$ and $A$ are labeled by three integers $m_\xi$, $m_\eta$, $m_{\chi}$, characterizing the properties of the solutions of the coupled system.
We will focus on bound-state solutions, for which $\bar{E}<0$.

It is then easily seen that this gravitational problem is completely equivalent to the quantum-mechanical Schr\"odinger equation
for the electronic energy of a one-electron heteronuclear diatomic molecule. In particular, if we identify $Z_1+Z_2=\mu\alpha$, $Z_2-Z_1=\mu\Delta\alpha$ then the equations above are mathematically equivalent to the quantum-mechanical problem (in atomic units), where the nuclei have atomic numbers $Z_i$ each and are separated by a fixed distance $D\equiv 2a$~\cite{Burrau:1927a,Wilson:1928,doi:10.1063/1.1726555,Nickel_2011,doi:10.1063/1.527130}.
In particular, this system also describes the ionized dihydrogen molecule~\cite{Burrau:1927a,Wilson:1928}.
We thus have a formal equivalence between two similar systems, that of a molecule governed by electromagnetism and 
a simple binary system in a Newtonian setting. We will see below that the inclusion of full general-relativistic effects alters this picture only slightly.

\begin{table}[htb]
  \caption{Eigenvalues for equal-mass binaries corresponding to $\alpha=0.2$, and to two different binary separations $2a=10M,\,60M$. The energy $\bar{E}$ and angular separation $A$ are labeled by three integers $(m_\xi, m_\eta, m_{\chi})$ in a manner analogous to the eigenvalues of the hydrogen atom in quantum mechanics. %All dimensionful quantities are normalized by the total mass $M$.
\label{table:eigen_spheroidal}}
\begin{ruledtabular}
\begin{tabular}{ccccc}
% \hline
%
 & \multicolumn{2}{c}{$a=5M$}         & \multicolumn{2}{c}{$a=30M$}\\
%  \hline
$(m_\xi, m_\eta, m_{\chi})$&$A$              &$10^2\times\bar{E}M $       & $A$      &$10^2\times\bar{E}M $\\
\hline\hline
(0,0,0)              &$-0.0129$        & $-0.386$      &$-0.342$  & $-0.272$\\%\hline
(1,0,0)              & $-0.00327$      & $-0.0981$     &$-0.0993$ & $-0.0817$\\%\hline
(2,0,0)               &$-0.00146$      & $-0.0439$    &$-0.0468$ & $-0.0387$\\%\hline
(0,2,0)              &$5.998$          & $-0.0445$    &$5.915$   & $-0.0453$\\%\hline
(1,2,0)              &$5.999$          & $-0.0250$    &$5.952$   & $-0.0254$ \\%\hline
(2,2,0)              &$5.999$          & $-0.0160$    &$5.970$  & $-0.0162$% \\%\hline
\end{tabular}
\end{ruledtabular}
\end{table}
Equations~\eqref{eq:spheroidals} are of spheroidal type. The first is an ``angular''-type scalar spheroidal equation~\cite{doi:10.1063/1.527130,Berti:2005gp}, and it is coupled to the second
(radial) equation through the (unknown) energy $\bar E$ and separation constant $A$.
We have solved this system to find the characteristic energies $\bar E$, with two different methods. We use direct integration of the ordinary differential equations, shooting to the energy; in addition, we used a high-accuracy continued fraction approach to solve the same problem~\cite{doi:10.1063/1.527130}. Our numerical results for selected values of the separation are shown in Table~\ref{table:eigen_spheroidal}.

%%%%%%%%%%%%%%%%%%%%%%%%%%%%%%%%%%%%%%%%%%%%%%%%%%%%%%%%%%%%%%%%%%%%%%%%%%%%%
\subsection{The single black hole limit\label{subsec:single}}
%%%%%%%%%%%%%%%%%%%%%%%%%%%%%%%%%%%%%%%%%%%%%%%%%%%%%%%%%%%%%%%%%%%%%%%%%%%%%

At zero separation, we are effectively dealing with one single BH, for which the energy levels are known to high precision.
In spherical polar coordinates $(\bar r,\,\bar\theta,\,\bar\varphi)$ the eigenfunction is
\begin{align}
\bar \psi &= e^{-i\bar{E}\bar t}\gamma^{\ell}\mathcal{Y}^m_{\ell}(\bar \theta,\bar \varphi)e^{-\gamma/2}L_n^{2\ell+1}(\gamma)\,,\label{eq:atom_limit_wave}\\
\gamma&\equiv \frac{2M\mu^2\bar r}{\ell+n+1}\,,
\end{align}
with $\mathcal{Y}^m_{\ell}$ being the scalar spherical harmonics and $L_n^{2\ell+1}$ a generalized Laguerre polynomial (it's a polynomial of order $n$ in its argument). Note that $M=M_1+M_2$ is the BHB mass.
With these definitions and conventions, the energy eigenvalue is
\be
\bar{E}=-\frac{\mu \alpha^2}{2(\ell+n+1)^2}\,,
\ee
up to ${\cal O}(\alpha^3)$.
Higher-order expansions in $\alpha$ can be calculated using well-known techniques and are shown in Ref.~\cite{Baumann:2019eav}, with a different state label.
Notice that we follow the state labeling of Ref.~\cite{Brito:2015oca}.
The scalar profile of mode $(n,\ell,m)$ decays spatially as $r^{\ell+n}e^{-\gamma/2}$ and the angular profile is dictated by the corresponding spherical harmonic.
The spatial extent of the scalar configuration is defined by the exponential decay, and is of order ${\cal S}\sim 1/(M\mu^{2})$.

Note that the prolate coordinates [Eq.~\eqref{eq:prolate-def}] are also given by
\be
\xi=\frac{r_1+r_2}{2a}\,,\qquad \eta=\frac{r_1-r_2}{2a}\,.
\ee
%
%where $r_i$ is the distance to BH ``i.'' 
Defining spherical coordinates such that
\begin{align*}
  \bar x & =\bar r\cos\bar \theta \,, \\
  \bar y & =\bar r\sin\bar \theta\sin\bar\varphi \,,\\
  \bar z & =\bar r\sin\bar \theta\cos\bar\varphi \,,
\end{align*}
one finds, when $a\to 0$,
\[
\xi  \to \frac{\bar r}{a}\,,\qquad \eta \to \cos \bar \theta \,,\qquad \chi =\bar \varphi\,.
\]
The relation between quantum numbers is then, in this limit~\cite{BATES196813}
\begin{equation}
  m_\xi=n \,, \qquad
  m_\eta=\ell-|m| \,, \qquad
  m_\chi=m \,.
  \label{correspondence_numbers}
\end{equation}

For small separations $a\mu$, our results---shown in Table~\ref{table:eigen_spheroidal}---for the fundamental ($n=0$) mode are compatible with (up to terms of order ${\cal O}(a^3\mu^3)$)
\begin{align}
\bar{E}&=-\frac{\mu\alpha^2}{2(\ell+1)^2}\nonumber\\
&\times\left(1+\frac{4a^2\mu^2\left(\ell(\ell+1)-3m_{\chi}^2\right)\left(\alpha^2-\Delta\alpha^2\right)}{\ell(\ell+1)^2(2\ell-1)(2\ell+1)(2\ell+3)}\right)\,.\label{eq:eigenvalue_expansion}
\end{align}
This analytic expansion was derived by Bethe, and in subsequent analytical work for the dihydrogen molecule and generalizations thereof~\cite{BETHE1933,doi:10.1063/1.1726555,Nickel_2011,BATES196813,article:abramov,PhysRevLett.52.1112}. Our own numerical results in the limit of small binary separation are in perfect agreement with such expression.

We note that for our numerical simulations shown in Sec.~\ref{sec:results}, we will use a frame rotated by 90 degrees, which therefore superposes different $m$ states.
In addition, and more importantly, at finite separation $a$, we can no longer use Eq.~\eqref{correspondence_numbers}: a given state in the $(m_\xi\,, m_\eta\,,m_\chi)$
basis involves, generically, a superposition of {\it all} modes in a spherical harmonic decomposition, such as that of Sec.~\ref{sec:results}.

%%%%%%%%%%%%%%%%%%%%%%%%%%%%%%%%%%%%%%%%%%%%%%%%%%
\subsection{Relation with classical closed orbits}
%%%%%%%%%%%%%%%%%%%%%%%%%%%%%%%%%%%%%%%%%%%%%%%%%%
%
Particle analogies of wave equations often play an important role in several
physical phenomena. In particular, a lot of BH physics described by wave
equations have particle descriptions which helps us to understand the system from a
different point of view. For example, BH quasinormal modes (QNMs) can be
related to closed null orbits around the BH based on the WKB approximation, where
the QNM frequencies can be estimated from the orbital periods~\cite{Cardoso:2008bp}.
Conversely, the existence of closed null orbits around relativistic objects may hint at the possibility of the existence of QNMs~\cite{Bernard:2019nkv}.

We will therefore study the particle analog of the system we have been
considering of a massive scalar field around a BHB.
As we will now see, this particle description can be derived from the WKB approximation of
Schr\"odinger's equation and can be applied to states around a general
separable spacetime.

In quantum mechanics, it is well known that the first order of the WKB approximation corresponds to the Hamilton-Jacobi equation for a classical particle, and
the energy spectrum can be related to the particle motion through the Bohr-Sommerfeld condition~\cite{weinberg_2015}\footnote{Strictly speaking, the Bohr-Sommerfeld condition is $\oint p_{k} dq_{k}=2\pi \left(n_{k} + \frac{1}{2}\right)$; however, since we are only interested in its large-$n_k$ limit, we have neglected the $\frac{1}{2}$ term in Eq.~\eqref{eq:Bohr-Sommerfeld condition}.}
\begin{equation}
\oint p_{k} dq_{k}=2\pi n_{k} \,,
\label{eq:Bohr-Sommerfeld condition}
\end{equation}
where $q_{k}$ and $p_{k}$ are the canonical coordinates and $n_{k}$ is an integer.
The integral is performed over a closed classical orbit which is described by the corresponding Hamilton-Jacobi equation.
Since this is derived from the WKB approximation, we can use this relation in the large-$n_{k}$ limit to estimate the frequency of the bound states of Sec.~\ref{sec:equivalence_molecule}.

Furthermore, from the classically allowed regions of the Hamilton-Jacobi equation, we can have an idea about the spatial profile of the wave function.
Since, as shown in the previous subsection,
our system can be described by Schr\"odinger's equation, 
we can apply these arguments to get the relation between (truly) bound states and closed orbits of a massive particle.\footnote{These are actual bound states because we are using a Newtonian approximation, and horizons are not present.}

Let us then consider the classical motion under Newton's
potential [Eq.~\eqref{eq:potential}], which corresponds to the motion of a massive
particle (of mass $\mu$) around a BHB, in the binary rest frame. The Lagrangian for the particle is
%
%\begin{widetext}
\[
\mathcal L(\bar x^i(\bar t),\partial_t \bar x^i( \bar t))=\frac{\mu}{2}\left(
  \partial_{ \bar  t}{ \bar x}( \bar t)^{2}
  +\partial_{ \bar  t}{ \bar y}( \bar t)^{2}
  +\partial_{ \bar t}{ \bar z}( \bar t)^{2}
\right)-V\,,
\]
%\end{widetext}
where $( \bar x( \bar t), \bar y( \bar t), \bar z( \bar t))$ are the particle's coordinates.
Using the coordinates $(\xi,\eta,\chi)$, the Hamilton-Jacobi equation reads
\begin{widetext}
\begin{equation}
  2\mu a^{2}\partial_{ \bar t}S
  + \frac{\left(\xi ^2-1\right) (\partial_{\xi}S)^2
    +\left(1-\eta ^2\right) (\partial_{\eta}S)^2
    -2 \mu ^2 a \left[ (M_1+M_2) \xi + (M_1-M_2) \eta \right]}{(\xi - \eta) (\xi + \eta)}
  +\frac{(\partial_{\chi}S)^2}{\left(1-\eta ^2\right) \left(\xi ^2-1\right)}=0\,,
  \label{eq:HJ}
\end{equation}
\end{widetext}
where $S(\bar t, \xi, \eta,\chi)$ is Hamilton's principal function.
In these coordinates the Hamilton-Jacobi equation is separable; we thus write
\begin{equation*}
  S(\bar t,\xi,\eta,\chi) =  S_{\xi}(\xi) + S_{\eta}(\eta)
  + m_{\chi}\chi - \bar E \bar{t} \,,
  % +S_{\chi}(\chi)+S_{t}(t)\,.
\end{equation*}
and after substituting into Eq.~\eqref{eq:HJ}, we obtain
%
% \begin{widetext}
\begin{equation}
\begin{aligned}
(S_{\eta}')^{2}&=
\frac{2a^{2}\mu}{1-\eta^{2}}\left(
C_{0}
-\bar E\eta^{2}+
\frac{\Delta\alpha}{a}\eta
-\frac{m_{\chi}^{2}}{2a^{2}\mu}\frac{1}{1-\eta^{2}}
\right)\,,\\
(S_{\xi}')^{2}&=\frac{2a^{2}\mu}{\xi^{2}-1}\left(
-C_{0}
+\bar E\xi^{2}+\frac{\alpha}{a}\xi-\frac{m_{\chi}^{2}}{2a^{2}\mu}\frac{1}{\xi^{2}-1}
\right)\,,
\end{aligned}
\label{eq:Set Sxi}
\end{equation}
where $C_{0}$ is a separation constant.
Comparing with Eq.~(\ref{eq:spheroidals}), we see that
$2a^{2}\mu C_{0}$ corresponds to $A$.
For classical motion, the right-hand side of Eq.~\eqref{eq:Set Sxi} must be positive, and the parameter space where this happens corresponds to the classically allowed region.
The topology of this region can change depending on the parameters; for simplicity, we will focus on equal-mass binaries ($\Delta\alpha=0$) and on
bound states around the binary ($\bar E<0$) with $m_{\chi}=0$.

\begin{figure*}[tb]
\includegraphics[width=0.3\textwidth]{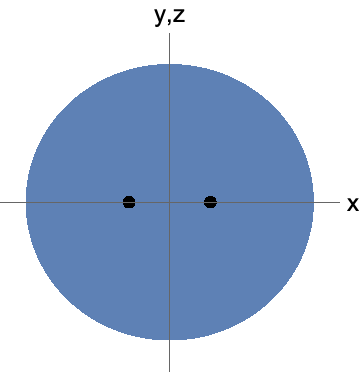} \quad
\includegraphics[width=0.3\textwidth]{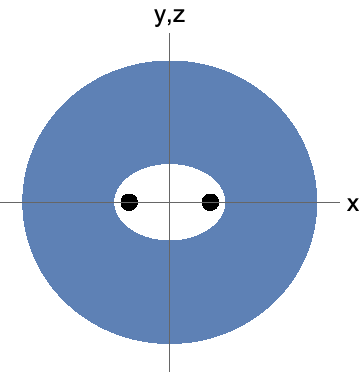} \quad
\includegraphics[width=0.3\textwidth]{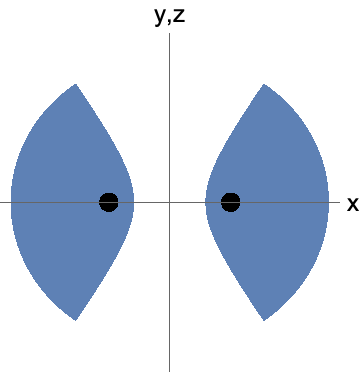}
\caption{Typical shape of the classically allowed region for a massive particle
  with $m_{\chi}=0$ around an equal-mass BHB. %, whose total mass is $M$. 
	%Parameters $(a/M,\mu M,\bar{E}_{0}M,C_{0}M)$ are
  %$(5,0.2,-0.01,0.01)$ (first panel), $(5,0.5,-0.02,0.1)$ (second panel), and
  %$(5,0.5,-0.02,-0.01)$ (third panel). 
	Black circles denote the BHs.
	The first panel shows the allowed region for $C_{0}>0$ and $|\bar E|-\alpha/a+C_{0}<0$,
	and the second panel shows the allowed region for $C_{0}>0$ and $|\bar E|-\alpha/a+C_{0}>0$.
	These describe orbits around the binary.
	The third panel stands for $C_{0}<0$ and $|\bar E|-\alpha/a+C_{0}<0$,
	and describes orbits around each individual BH.
\label{Classical_allowed_region1}}
\end{figure*}

Considering first the $C_{0}>0$  case, we see
% In this case, the allowed region for $\eta$ is $[-1,1]$.
from Eq.~\eqref{eq:Set Sxi} that $\eta$ is unconstrained ($-1 \le \eta \le 1$),
and the allowed region for $\xi$ is determined from
\begin{equation}
  |\bar E|\xi^{2}-\frac{\alpha}{a}\xi+C_{0}<0 \,.
  \label{eq:allowed-region}
\end{equation}
When $|\bar E|-\frac{\alpha}{a}+C_{0}<0$, the allowed region for $\xi$ is $\xi \in [1,\xi_{+}]$, and
when  $|\bar E|-\frac{\alpha}{a}+C_{0}>0$, the allowed region is $\xi \in [\xi_{-},\xi_{+}]$.
Here, $\xi_{\pm}$ are solutions of the equation $|\bar E|\xi^{2}-\frac{\alpha}{a}\xi+C_{0}=0$:
% that is
%
\begin{equation}
\xi_{\pm}=\frac{1}{|\bar E|}\left(
\frac{\alpha}{2a}\pm\sqrt{\frac{\alpha^{2}}{4a^{2}}-|\bar E|C_{0}}
\right) \,.
\end{equation}
Therefore, when $C_{0}>0$,
the particle motion is an orbit around the binary (see the first and second panels in Fig.~\ref{Classical_allowed_region1}).

We now focus on the $C_{0}<0$ case.
The allowed region for $\eta$ is
% $1>\eta>\sqrt{\frac{|C_{0}|}{|E|}}$ or $-\sqrt{\frac{|C_{0}|}{|E|}}>\eta>-1$,
$\eta \in \left[-1,-\sqrt{\frac{|C_{0}|}{|\bar E|}}\right] \cup \left[\sqrt{\frac{|C_{0}|}{|\bar E|}}, 1\right]$ for $|C_{0}|<|\bar E|$.
Existence of solutions to Eq.~\eqref{eq:allowed-region} then implies that $|\bar E|-\frac{\alpha}{a}-|C_{0}|<0$, and the corresponding allowed region for $\xi$ is then $[1,\xi_{+}]$ (see the third panel in Fig.~\ref{Classical_allowed_region1}).
In this case, the motion is an orbit around each individual BH.

Since the WKB approximation is valid only in the high-frequency limit,
we cannot easily apply this classical picture to the states from the previous subsection. We can, however, see that there are two distinct profiles for these states: one corresponding to configurations bound to each individual BH, and 
another corresponding to a configuration bound to the binary as a whole.

In order to compare the particle picture to the states of the previous subsection, 
let us discuss the spectrum of these states using the Bohr-Sommerfeld condition Eq.~\eqref{eq:Bohr-Sommerfeld condition}.
From this condition we immediately see that $m_{\chi}$ must be the integer $n_{\chi}$ and we further get
\begin{widetext}
\begin{subequations}
\begin{align}
\int_{\eta_{-}}^{\eta_{+}}\sqrt{
\frac{2a^{2}\mu}{1-\eta^{2}}\left(
C_{0}
-\bar E\eta^{2}+
\frac{\Delta\alpha}{a}\eta
-\frac{m_{\chi}^{2}}{2a^{2}\mu}\frac{1}{1-\eta^{2}}
  \right)}d\eta&=2\pi n_{\eta}\,,\\
\int_{\xi_{-}}^{\xi_{+}}\sqrt{
\frac{2a^{2}\mu}{\xi^{2}-1}\left(
-C_{0}
+\bar E\xi^{2}+\frac{\alpha}{a}\xi-\frac{m_{\chi}^{2}}{2a^{2}\mu}\frac{1}{\xi^{2}-1}
\right)}d\xi&=2\pi n_{\xi}\,,
\end{align}
\label{eq:molecule Bohr-Sommerfeld condition}
\end{subequations}
\end{widetext}
where $n_{\xi}$ and $n_{\eta}$ are integers, and the integration limits $\eta_{\pm}$ and $\xi_{\pm}$ are the roots of the expressions inside each square root.
Since Eqs.~\eqref{eq:molecule Bohr-Sommerfeld condition} cannot be integrated analytically let us consider the hydrogen atom limit ($a\to 0$) by fixing $\ell^{2}=2a^{2}\mu C_{0}$ and $r=a\xi$ in the equal-mass case ($\Delta \alpha=0$).
We obtain the spectrum
\begin{align*}
\ell-|m_{\chi}|&=n_{\eta}\,,\\
-\alpha\sqrt{\frac{\mu}{2|\bar E|}}-\ell&=n_{\xi}\,.
\end{align*}
These expressions are in good agreement with the spectrum of the hydrogen atom
in the large-$n_{\eta}$ and large-$n_{\xi}$ limit, confirming the validity of this particle picture.

%%%%%%%%%%%%%%%%%%%%%%%%%%%%%%%%%%%%%%%%%%%%%%%%%%%%%%%%%%%%%%%%%%%%%%%%%%%%%
\subsection{Corrections induced by orbital motion}
\label{Motion_binary_effects}
%%%%%%%%%%%%%%%%%%%%%%%%%%%%%%%%%%%%%%%%%%%%%%%%%%%%%%%%%%%%%%%%%%%%%%%%%%%%%

Having solved Eq.~(\ref{eq:schrod-pert}) to zeroth order in $\Omega$, let
us now consider, perturbatively, the effect of rotation. The first-order
correction in $\Omega$ to the unperturbed energy levels that we have just
computed is given by the expectation value of the rotation operator
$L_{\bar z} = -i \partial_{\bar \varphi}$ for the system in the unperturbed
state. In the coordinates $(\xi,\eta,\chi )$ this operator takes the form
\begin{equation}
\label{eq:Lz}
L_{\bar z} = \frac{i \bar y}{a(\xi^2-\eta^2)}
            \left( \xi \partial_{\eta} + \eta \partial_{\xi} \right) +
            \frac{i \bar z \, \xi \eta}{a(\xi^2-1)(1-\eta^2)} \partial_{\chi} \,.
\end{equation}
Since $\bar z \sim \cos \chi$ and $\bar y \sim \sin \chi$, we can easily see
that the expectation value of this operator in an
eigenstate [Eq.~(\ref{eq:eingenstate})] is zero,
$\langle L_{\bar z} \rangle_{\bar \psi} = 0$. This shows that rotation effects only
manifest themselves at order $\Omega^2$ and will therefore be
neglected.

%%%%%%%%%%%%%%%%%%%%%%%%%%%%%%%%%%%%%%%%%%%%%%%%%%%%%%%%%%%%%%%%%%%%%%%%%%%%%
\section{Setup for time evolutions}
\label{sec:numerics}
%%%%%%%%%%%%%%%%%%%%%%%%%%%%%%%%%%%%%%%%%%%%%%%%%%%%%%%%%%%%%%%%%%%%%%%%%%%%%

In the previous section we discussed, in a perturbative setup, the nonrelativistic limit of a massive
scalar field around a BHB and found molecule-like states which were labeled therein with three parameters $(m_{\xi},m_{\eta},m_{\chi})$. We
will now numerically solve the Klein-Gordon equation around a BHB to
construct these (quasi)bound states through time evolutions.

%%%%%%%%%%%%%%%%%%%%%%%%%%%%%%%%%%%%%%%%%%%%%%%%%%%%%%%%%%%%%%%%%%%%%%%%%%%%% 
\subsection{Numerical implementation}
\label{sec:implementation}
%%%%%%%%%%%%%%%%%%%%%%%%%%%%%%%%%%%%%%%%%%%%%%%%%%%%%%%%%%%%%%%%%%%%%%%%%%%%%

For our numerical implementation we ignore the backreaction of the massive
scalar field on the BHB spacetime and follow the approach described
in Ref.~\cite{Bernard:2019nkv}. In this approach one builds an approximate BHB
background metric using the construction outlined in detail in Mundim
et~al.~\cite{Mundim:2013vca} (see also
Ref.~\cite{Johnson_McDaniel_2009} for the equivalent construction used in the
context of generating BHB initial data). It is important to
mention that for our present construction we turn off the emission of gravitational
radiation and therefore always consider binaries with constant separation. We do {\it not} turn off scalar radiation; the Klein-Gordon equation is evolved in full generality.

Our approach then consists in using this approximate BHB metric construction and
solving the Klein-Gordon equation~(\ref{eq:KG}) in this spacetime. We emphasize
that this metric, even though it is time-dependent, is \emph{prescribed}---that is,
it is not time-evolved. Our task is then to numerically solve Eq.~(\ref{eq:KG})
on a time-dependent background. To do so we write the equation in a first-order
form by introducing
\begin{equation}
K_{\phi}\equiv-\frac{1}{2N} \left(
\partial_{t}-\mathcal{L}_{\beta}
\right)\phi \,,
\label{eq:Kphi}
\end{equation}
where $N$ and $\beta^{i}$ are the lapse function and shift vector, respectively.
The resulting system is numerically evolved with a method of lines approach.

For our numerical evolutions, we use the Einstein~Toolkit
infrastructure~\cite{Loffler:2011ay,Zilhao:2013hia,EinsteinToolkit:2019_10} with
Carpet~\cite{Schnetter:2003rb,CarpetCode:web} for mesh refinement capabilities
and the multipatch infrastructure Llama~\cite{Pollney_2011}. The scalar field
equations are evolved in time by adapting the ScalarEvolve code available
in Ref.~\cite{Canuda_2020_3565475}, which was first used and described
in Ref.~\cite{Cunha:2017wao}. Since the background metric uses harmonic coordinates,
in our evolutions we further excise the BH interior with the procedure outlined
in Ref.~\cite{Assumpcao:2018bka}. The overall infrastructure and evolution are
essentially the same those used and tested in Ref.~\cite{Bernard:2019nkv}.
We employ fourth-order-accurate finite-differencing stencils to approximate spatial derivatives, but there are lower-order elements in the code---in particular the so-called prolongation operation is only second-order accurate in time. Our results are compatible with a convergence order between orders 2 and 3, which is consistent with the overall setup. See Appendix~\ref{sec:Convergence test} for further details.

%%%%%%%%%%%%%%%%%%%%%%%%%%%%%%%%%%%%%%%%%%%%%%%%%%%%%%%%%%%%%%%%%%%%%%%%%%%%%
\subsection{Initial data}
%%%%%%%%%%%%%%%%%%%%%%%%%%%%%%%%%%%%%%%%%%%%%%%%%%%%%%%%%%%%%%%%%%%%%%%%%%%%%

We will evolve two different types of scalar field initial data, which will be
referred to as ``nonspinning'' and ``spinning'' initial data. The first of
these consists of a momentarily static Gaussian profile given by
\begin{equation}
\phi = Ae^{-r^2/(2\sigma^{2})}\,, \qquad
K_{\phi} = 0\,,
\label{eq:static_ID}
\end{equation}
where $A$ and $\sigma$ denote the amplitude and width of the Gaussian pulse, respectively.

Secondly, we will evolve configurations which, unlike the previous construction, have angular momentum. These are the ``spinning'' initial data, for which
\begin{equation}
\phi = R(r)\mathcal{A}(t,\theta,\varphi)\,,  \label{eq:clump initial data}
\end{equation}
with
\begin{align}
R(r) &=  \frac{r}{\sigma}e^{-\frac{r}{2\sigma}}\,, \\
\mathcal{A}(t,\theta,\varphi) & = A_{1,1}\frac{1}{2}\sqrt{\frac{3}{2\pi}}\sin\theta\cos(\varphi+\omega t)\nonumber\\
&\quad+A_{1,-1}\frac{1}{2}\sqrt{\frac{3}{2\pi}}\sin\theta\cos(-\varphi+\omega t)\,,
\end{align}
and the initial configuration for the $K_{\phi}$ field can be trivially obtained from Eq.~\eqref{eq:Kphi}.
Here, $A_{l,m}$ is the amplitude of each $(l,m)$ mode, which can be freely specified,
and $\sigma$ is a typical width of the clump.
$(r,\theta,\varphi)$ are the standard spherical coordinates around the center of mass of the system.
The time dependence $m\varphi+\omega t~(m=\pm 1)$ in each term introduces angular momentum on the $z$ axis.
Initial data with negative $m$ are corotating with the BHB, while initial data with positive $m$ are antirotating.

%%%%%%%%%%%%%%%%%%%%%%%%%%%%%%%%%%%%%%%%%%%%%%%%%%%%%%%%%%%%%%%%%%%%%%%%%%%%
\subsection{Frequency extraction\label{subsec:rotating_frame_wave}}
%%%%%%%%%%%%%%%%%%%%%%%%%%%%%%%%%%%%%%%%%%%%%%%%%%%%%%%%%%%%%%%%%%%%%%%%%%%%%
%
To analyze our results, we decompose the evolved field at fixed radial distance into spherical
harmonics $Y^m_l(\theta, \varphi)$ as follows
\begin{equation}
  \phi(t,r=r_{\rm ext},\theta,\varphi) = \sum_{l=0}^{\infty} \sum_{m=-l}^l
  Y^m_l(\theta, \varphi) \phi_{lm}(t) \,.
  \label{eq:spherharm}
\end{equation}
Note that, since $\phi$ is real, $\phi_{lm}(t) = (-)^m \phi_{l,-m}^{\ast}(t)$ where $\ast$ denotes complex conjugation.
We will further Fourier-transform $\phi_{lm}(t)$ to check its frequency spectra
and compare with the results obtained in Sec.~\ref{sec:equivalence_molecule}. In
order to do so, however, we must note that the frequencies computed in
Sec.~\ref{sec:equivalence_molecule} were computed in the rest frame of the
binary, whereas here we will be extracting the data in the lab frame. To compare
the data we then need to change frames once again, which can be done as follows:

The Fourier spectra in the lab frame is computed through [$\phi=\phi(t,r_{\rm ext},\theta,\varphi)$]
\begin{align}
\mathcal{F}[\phi_{lm}](\omega) &= \int dt \, e^{i \omega t} \phi_{lm}(t)\nonumber\\
&= \int dt \, e^{i \omega t} \sin \theta \, d\theta d\varphi \, 
  Y^m_l{}^{\ast}(\theta, \varphi) \phi \,.
\end{align}
Given that $Y^m_l(\theta, \varphi) = N e^{i m \varphi} P^m_l(\cos \theta)$, where
$P^m_l$ are the Legendre polynomials and $N$ is the normalization constant,
using Eq.~\eqref{eq:torestframe}, we can write
%
% \begin{widetext}
\begin{align}
  \mathcal{F}[\phi_{lm}](\omega) & = \int d\bar t \, \sin \bar \theta \, d\bar \theta \,d\bar \varphi \,e^{i \bar t(\omega - m \Omega)} \nonumber\\
& \qquad \times N e^{-i m \bar \varphi}  P^m_l(\cos \bar \theta) \, \phi(\bar t,r_{\rm ext},\bar \theta, \bar \varphi + \Omega \bar t) \nonumber \\
                                 &=  \int d\bar t \, \sin \bar \theta d\bar \theta \,d\bar \varphi \,e^{i \bar t(\omega - m \Omega)} \notag \\
  & \qquad \times Y^m_l{}^{\ast}(\bar \theta,\bar \varphi) \,
    \bar \phi(\bar t,r_{\rm ext},\bar \theta,\bar \varphi) \nonumber \\
&= \mathcal{F}[\bar \phi_{lm}](\omega - m \Omega) \,,\label{eq:Fphi_Fbarphi}
\end{align}
% \end{widetext}
relating the frequencies computed in the lab frame to the ones computed in the
comoving frame.

Since in the following section we will be focusing on the real part of the multipolar
components let us also write
%
% \begin{widetext}
\begin{align}
\mathcal{F}_{\Re} & = \int dt \, e^{i \omega t} {\Re}(\phi_{lm}(t))\nonumber\\
&= \frac 1{2} \int dt \, e^{i \omega t} \left( \phi_{lm}(t) + \phi_{lm}^{\ast}(t) \right) \nonumber\\
&=\frac{1}{2}\left( \mathcal{F}[\phi_{lm}](\omega)+ (-)^{m} \mathcal{F}[\phi_{l,-m}](\omega)\right) \nonumber \\
                  &=\frac{1}{2}\big( \mathcal{F}[\bar \phi_{lm}](\omega - m \Omega) \nonumber \\
  & \qquad + (-)^{m} \mathcal{F}[\bar \phi_{l,-m}](\omega + m \Omega) \big)
\,, \label{eq:FRephi_Fbarphi}
\end{align}
where $\mathcal{F}_{\Re}\equiv \mathcal{F}[\Re (\phi_{lm})](\omega)$ and Eq.~\eqref{eq:Fphi_Fbarphi} is used in the last step. To connect with our upcoming
results we further need to take the real part of
Eq.~\eqref{eq:FRephi_Fbarphi},
\begin{align*}
  \Re\left(\mathcal{F}_{\Re}\right) & = \frac 1{4} \big(
 \mathcal{F}[\bar \phi_{lm}](\omega - m \Omega) 
+  (-)^m \mathcal{F}[\bar \phi_{l,-m}](\omega + m \Omega) \notag \\
& \qquad +  (-)^m \mathcal{F}[\bar \phi_{l,-m}](-\omega + m \Omega) \\
& \qquad +  \mathcal{F}[\bar \phi_{lm}](-\omega - m \Omega)
\big)
\end{align*}
% 
% \end{widetext}
where we have used
\[
  \mathcal{F}[\phi_{lm}^{\ast}](\omega) =
    (-)^m \mathcal{F}[\phi_{l,-m}](-\omega) \,.
\]
In conclusion, in the lab frame we expect to see a superposition of $\omega \pm m\Omega$ frequencies for each $m$ mode of the corotating frame.

%%%%%%%%%%%%%%%%%%%%%%%%%%%%%%%%%%%%%%%%%%%%%%%%%%%%%%%%%%%%%%%%%%%%%%%%%%%%%
\section{Gravitational molecules}
\label{sec:results}
%%%%%%%%%%%%%%%%%%%%%%%%%%%%%%%%%%%%%%%%%%%%%%%%%%%%%%%%%%%%%%%%%%%%%%%%%%%%%

We are now in a position to discuss the evolution of scalar fields on a background describing realistic BHBs.
We have evolved a battery of different configurations with the initial data of
Eqs.~(\ref{eq:static_ID}) and~(\ref{eq:clump initial data})
% We focus on equal-mass binaries, $M_{1}=M_{2}=M/2$.
on background spacetimes described by equal-mass binaries, $M_{1}=M_{2}=M/2$, for different values of binary separation $D$.
In this section we
will report on the results obtained with a subset of these runs, summarized in Tables~\ref{table:simulations1} and~\ref{table:simulations2}.
The numerical convergence for the simulations \texttt{nonspin2} is discussed in Appendix~\ref{sec:Convergence test}.
These results, as described below, are in good agreement with the nonrelativistic analysis of Sec.~\ref{section:non_relativistic}, and provide strong evidence for
the existence and formation of gravitational molecules with scalar fields.

\begin{table}[htb]
  \caption{List of simulations analyzed for the momentarily static Gaussian initial data of Eq.~\eqref{eq:static_ID}. Note that we have run a much larger set of simulations---listed here are only those that are analyzed later on.
\label{table:simulations1}}
\begin{ruledtabular}
\begin{tabular}{lccc}
 % \hline
 % \hline
Run          & $D/M$  & $\mu M$ & $\sigma /M$ \\
\hline
\texttt{nonspin1} &  60   &  0.5    &    12  \\
%\hline
\texttt{nonspin2} & 10    &  0.2    &    25 \\
%\hline
\texttt{nonspin3} & 60    & 0.2     &    25 \\
% \hline\hline
\end{tabular}
\end{ruledtabular}
\end{table}

\begin{table}[htb]
  \caption{List of simulations analyzed for the spinning initial data of Eq.~\eqref{eq:clump initial data}. \label{table:simulations2}}

\begin{ruledtabular}
\begin{tabular}{lcccccc}
 % \hline
 % \hline
Run   &         $D/M$ & $\mu M$  & $\sigma /M$ & $A_{1,1}$ &$A_{1,-1}$ & $\omega M$\\
\hline
\texttt{spin1} &  10  &  0.2     &  25         &  0        &  1       &   0.2 \\
% \hline
\texttt{spin2} & 60   & 0.2      & 25          & 0         & 1        & 0.2 \\
% \hline
\texttt{spin3} & 60   & 0.2      & 25          & 1         & 0        & 0.2 \\
% \hline\hline
\end{tabular}
\end{ruledtabular}
\end{table}

It is useful to keep in mind the different length scales involved in this problem.
As we saw in Sec.~\ref{subsec:single}, the scale ${\cal S}$ of a state around an isolated BH of mass $M_i$ is of order ${\cal S}_i\sim 1/(M_i\mu^{2})$ in the small-$M_i\mu$ limit.
In a binary of component masses $M_{1},\,M_{2}$, we thus have scales ${\cal S}_i$, and a global scale
${\cal S}_{\rm BHB}\sim \mu^{-2}/M$ where the total BHB mass is $M=M_{1}+M_{2}$. 
If ${\cal S}_{i}$ is much smaller than $D=2a$, the quasibound state can be formed around each BH and feels a tidal force from the companion object.
On the other hand, if ${\cal S}_{i}$ is much larger than $D$, the companion BH strongly disturbs such a state, destroying it.
However, as discussed in Sec.~\ref{section:non_relativistic}, we can expect that a quasibound state forms around the BHB.
Furthermore, if this state around the binary is stable, it should be formed starting from generic initial conditions.

Note also that timescales are important. A light-crossing timescale is of order $D/M$, whereas an orbital timescale is of order $2\pi \sqrt{D^{3}/M}\sim 200 M$ or $3000M$ for binaries separated by $D=10M$ or $60M$, respectively.

\begin{figure*}[thp]
\includegraphics[width=0.45\textwidth]{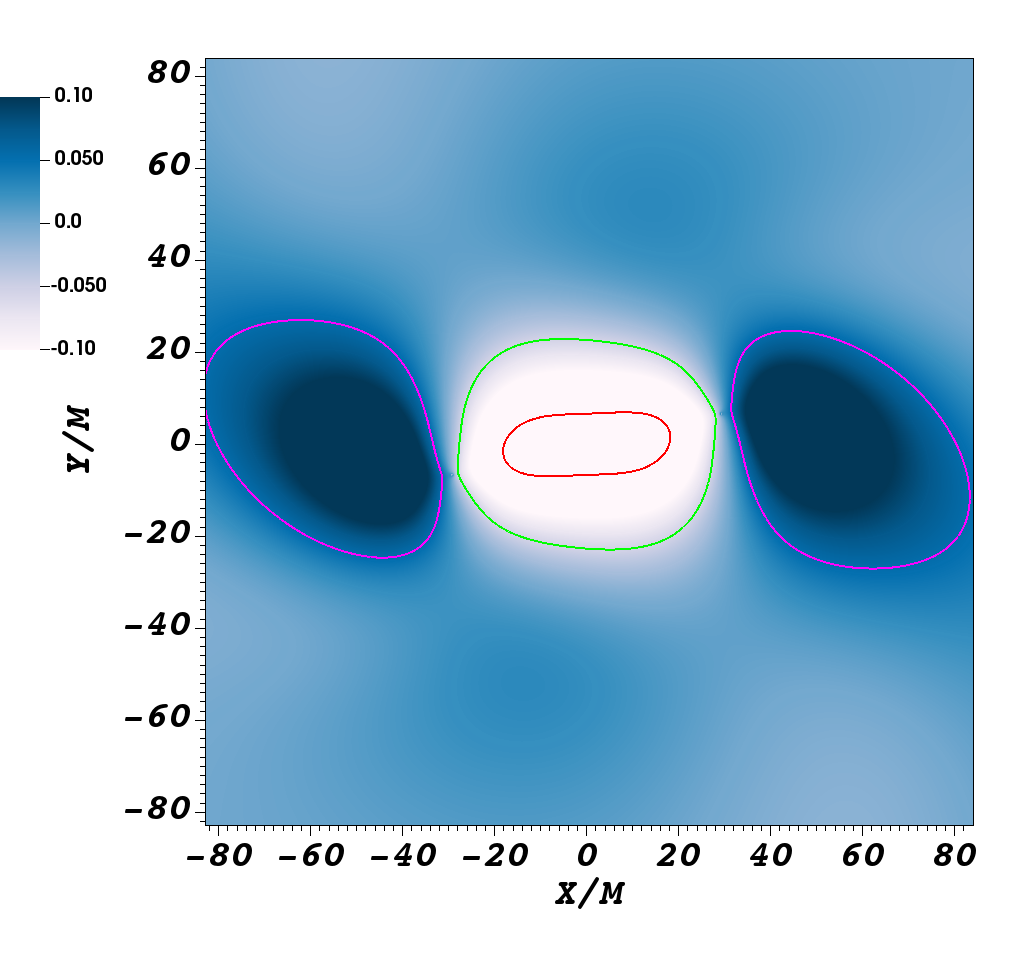}
  \quad
\includegraphics[width=0.45\textwidth]{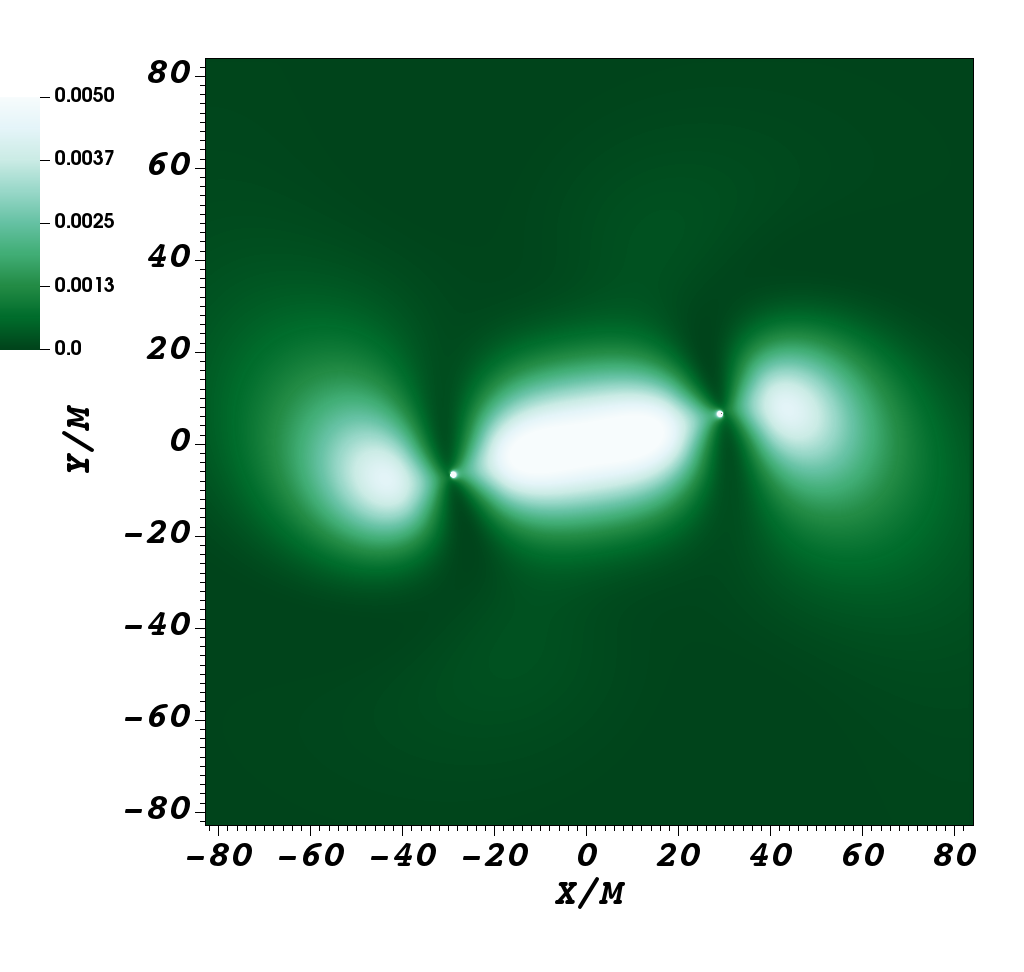}
\caption{Profile of a massive scalar field around an equal-mass BHB of total mass $M$.  The field has a dipolar, $l=1,\,m=0$ profile around each BH, which was obtained from evolving configuration \texttt{nonspin1} in Table~\ref{table:simulations1}.
The left panel shows the contours of a constant scalar field. 
Purple, green, and red lines represent lines of the constant scalar field which are $0.05$, $-0.05$, and $-0.16$, respectively.
Note that since the Klein-Gordon equation is linear in the scalar field and there is no backreaction on the metric, only their relative values are meaningful.
Colormaps are therefore unimportant to interpreting the results and will be omitted henceforth.
The right panel shows a contour plot of (scalar field) energy density.
The snapshots were taken at $t=1600M$.
\label{snapshot_scalarfield_t1600_M05_M05_D60_mu05_gaussian1_sigma12_Ac00_1_resolution1_SBP_v2}
}
\end{figure*}

%%%%%%%%%%%%%%%%%%%%%%%%%%%%%%%%%%%%%%%%%%%%%%%%%%%%%%%%%%%%%%%%%%%%%%%%%%%%%%%%%%%%%%%
\subsection{quasibound states around individual BHs\label{sec:Bound states around each BH}}
%%%%%%%%%%%%%%%%%%%%%%%%%%%%%%%%%%%%%%%%%%%%%%%%%%%%%%%%%%%%%%%%%%%%%%%%%%%%%%%%%%%%%%%

Based on the scales above, we expect that for large enough couplings $M\mu$ the scale ${\cal S}_i$ obeys ${\cal S}_i<D$, in which case the scalar cloud localizes around each BH but not around the binary.
In other words, we expect that BHs sufficiently far apart can support clouds as if in isolation. To test this, we evolve momentarily static Gaussian initial data, Eq.~\eqref{eq:static_ID}, corresponding to configuration \texttt{nonspin1} in Table~\ref{table:simulations1}.
For these parameters, the typical size of the cloud around each BH is ${\cal S}_{1}={\cal S}_{2}\sim 8M$, which is smaller than the BH separation.
We therefore expect that quasibound states form around each BH and our results confirm this, as shown in Fig.~\ref{snapshot_scalarfield_t1600_M05_M05_D60_mu05_gaussian1_sigma12_Ac00_1_resolution1_SBP_v2}. As can be seen, localized structures are apparent with the dominant mode being an $l=1,m=0$ state around each individual BH. Our results show that these configurations remain localized to the binary with negligible variation in its shape and topology for thousands of dynamical timescales, with only a slight variation in amplitude, hence qualifying as true quasibound states.

%%%%%%%%%%%%%%%%%%%%%%%%%%%%%%%%%%%%%%%%%%%%%%%%%%%%%%%%%%%%%%%%%%%%%%%%%%
\subsection{Global quasibound states from evolution of static initial data}
\label{sec:Momentarily_static_Gaussian_profile}
%%%%%%%%%%%%%%%%%%%%%%%%%%%%%%%%%%%%%%%%%%%%%%%%%%%%%%%%%%%%%%%%%%%%%%%%%%
We will now discuss molecular-like structures---i.e., scalar clouds around BHBs.
We focus on BH separations $D=10M,\,60M$, and we fix the mass of the scalar
field to $\mu M =0.2$. This corresponds to configurations \texttt{nonspin2} and \texttt{nonspin3}, respectively.
The scales are such that now the size of the quasibound states, if present, would encompass the binary when $D=10M$. We will see that even for $D=60M$
such global states exist.

\begin{figure*}[tbph]
\includegraphics[clip, width=3.5cm]{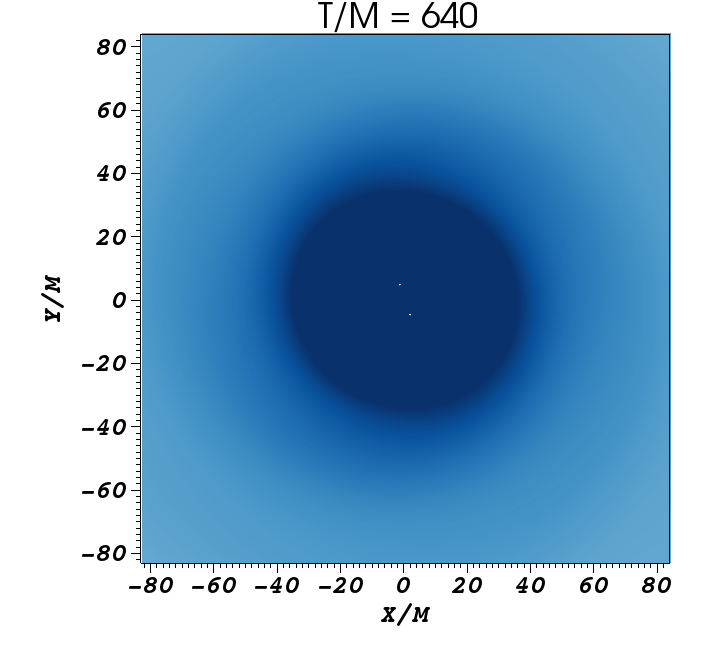}
\includegraphics[clip, width=3.5cm]{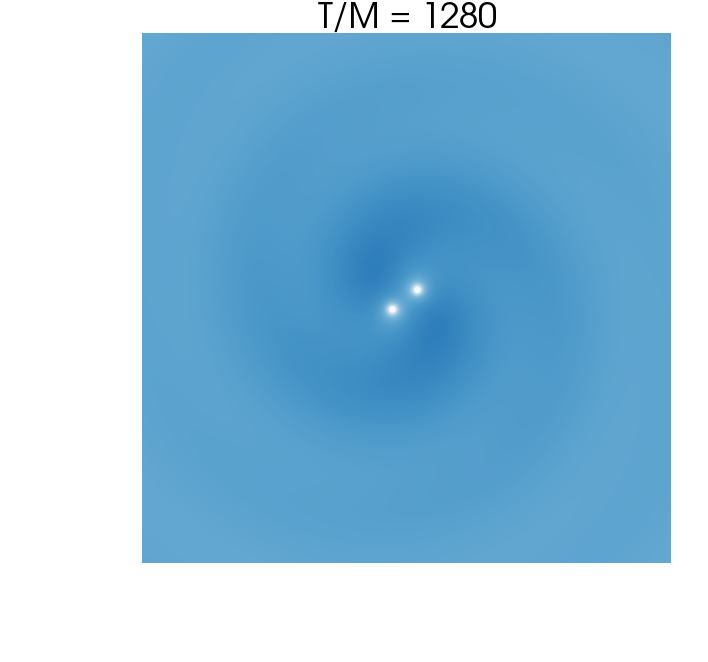}
\includegraphics[clip, width=3.5cm]{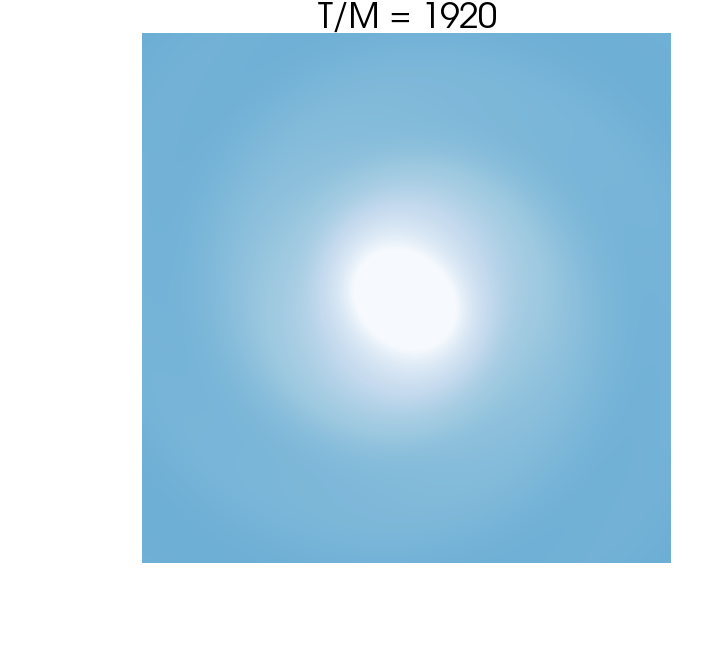}
\includegraphics[clip, width=3.5cm]{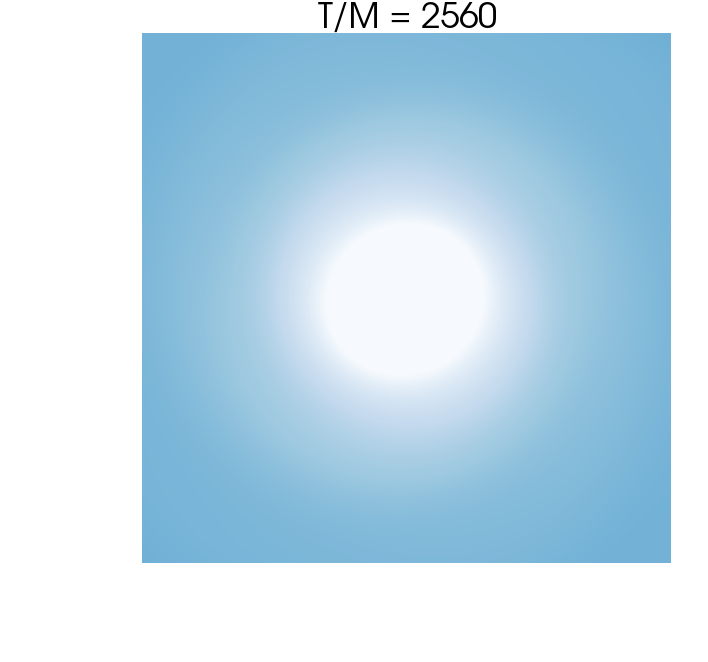}
\includegraphics[clip, width=3.5cm]{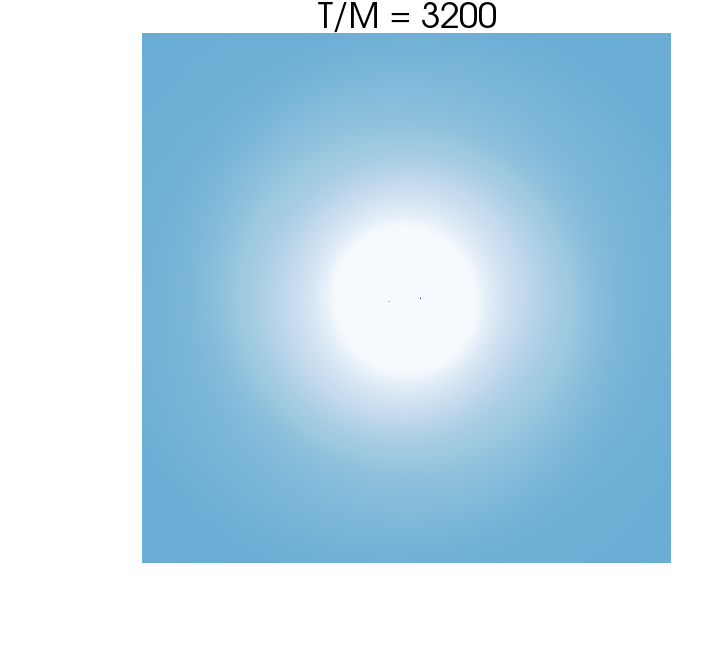}\\
\includegraphics[clip, width=3.5cm]{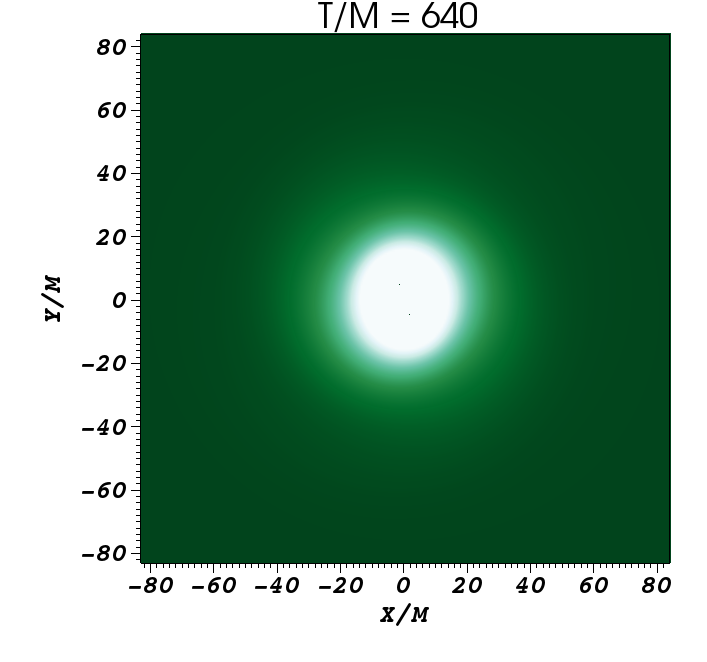}
\includegraphics[clip, width=3.5cm]{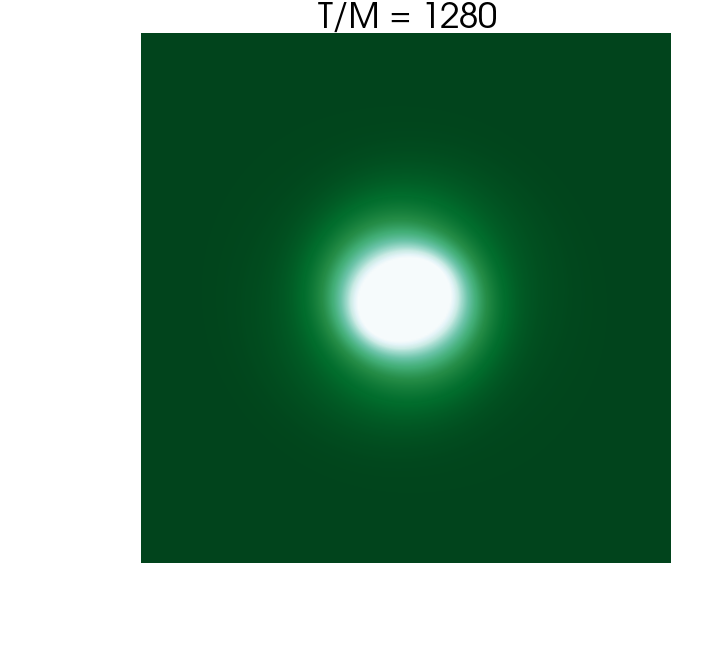}
\includegraphics[clip, width=3.5cm]{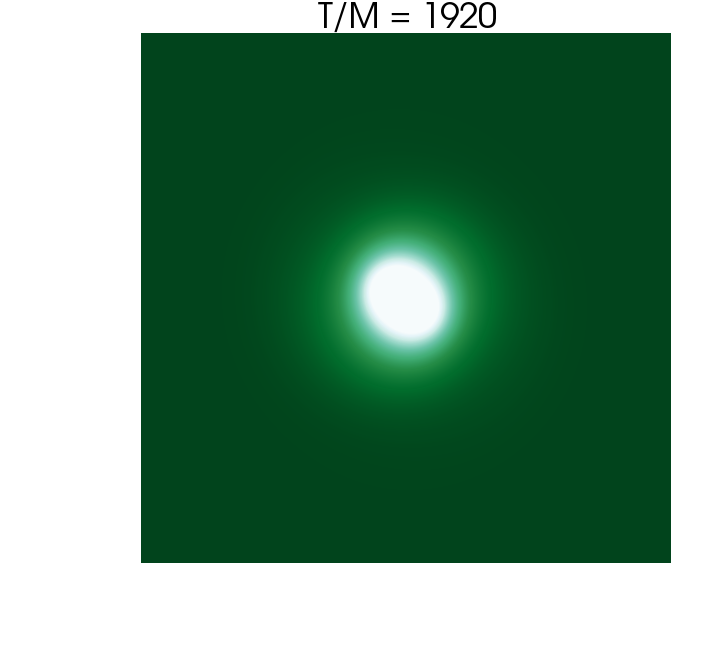}
\includegraphics[clip, width=3.5cm]{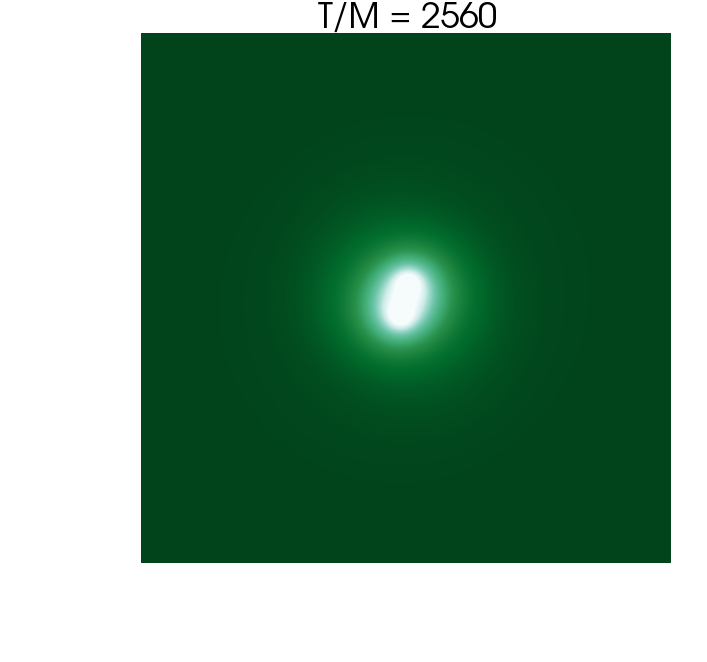}
\includegraphics[clip, width=3.5cm]{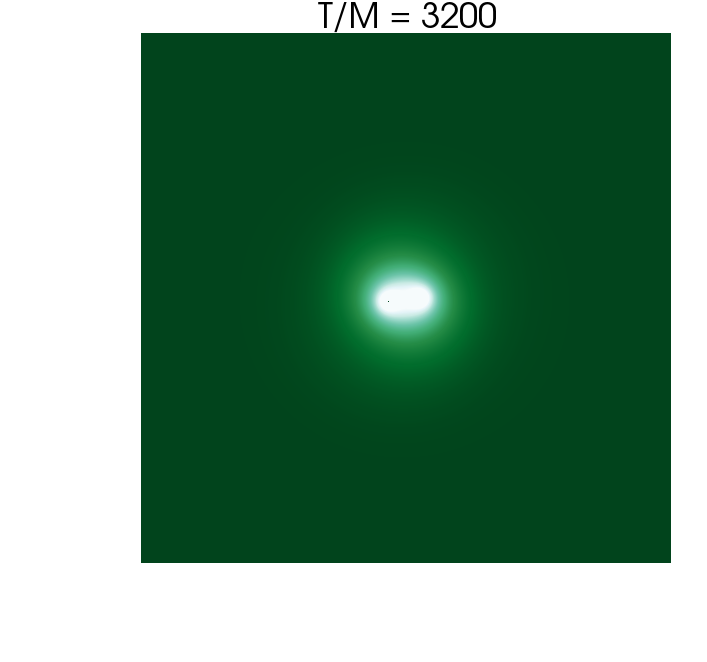}\\
\includegraphics[clip, width=3.5cm]{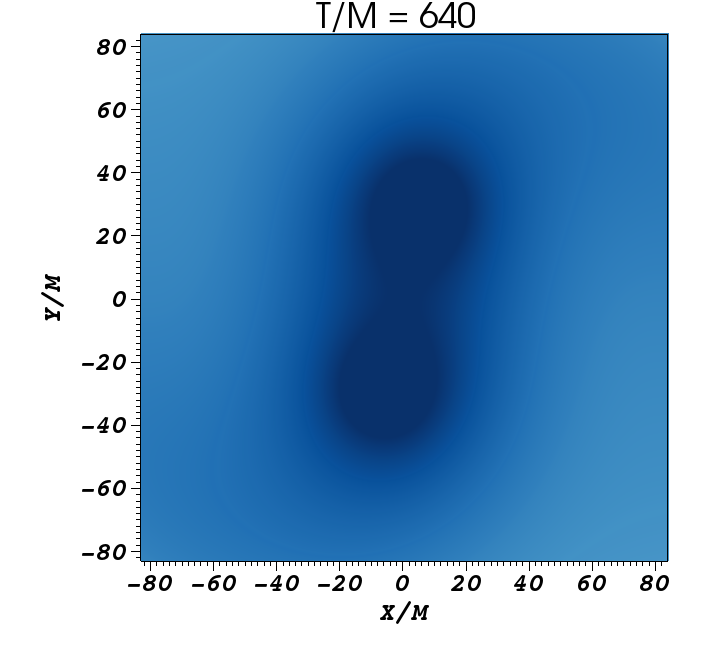}
\includegraphics[clip, width=3.5cm]{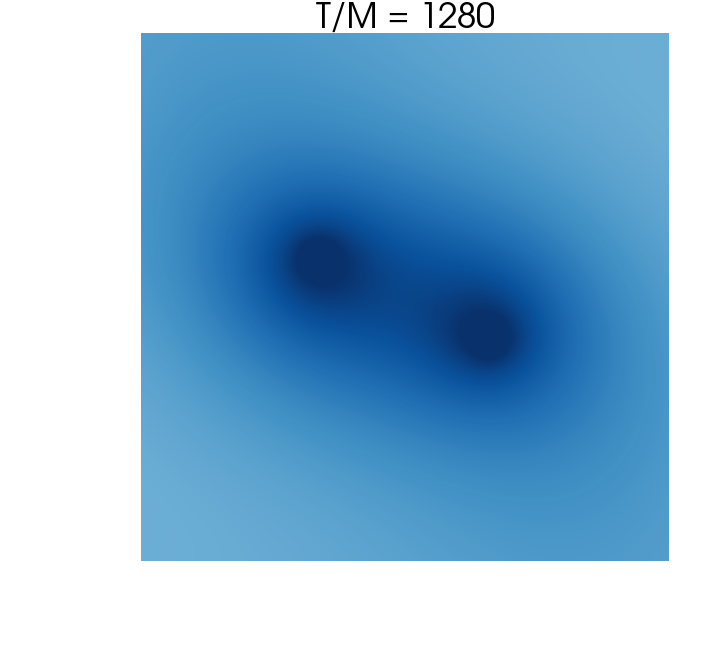}
\includegraphics[clip, width=3.5cm]{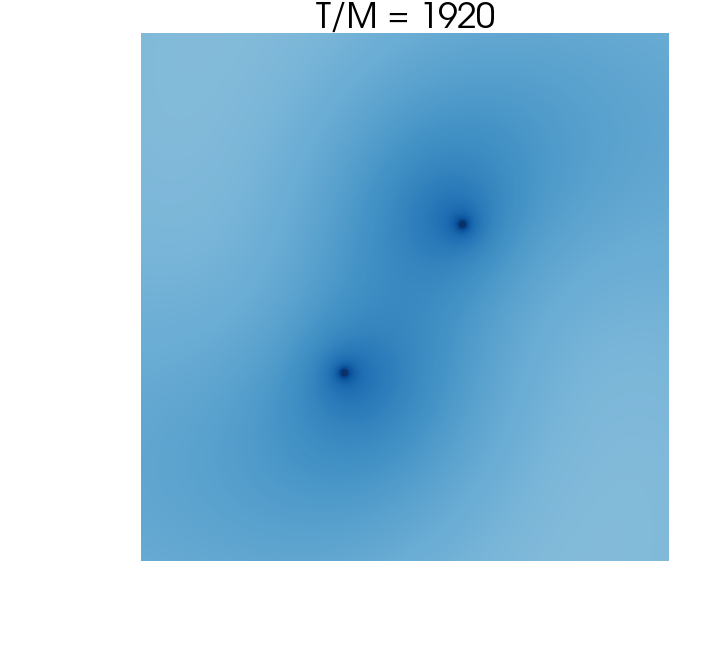}
\includegraphics[clip, width=3.5cm]{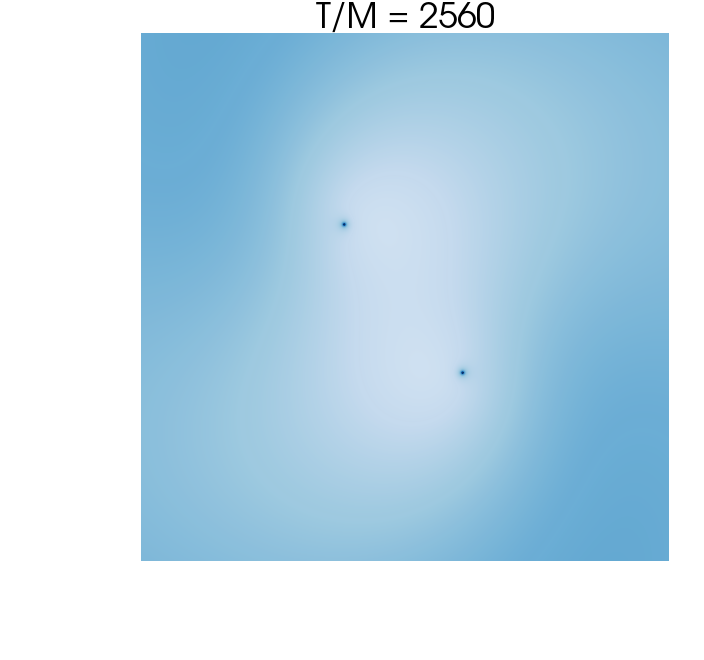}
\includegraphics[clip, width=3.5cm]{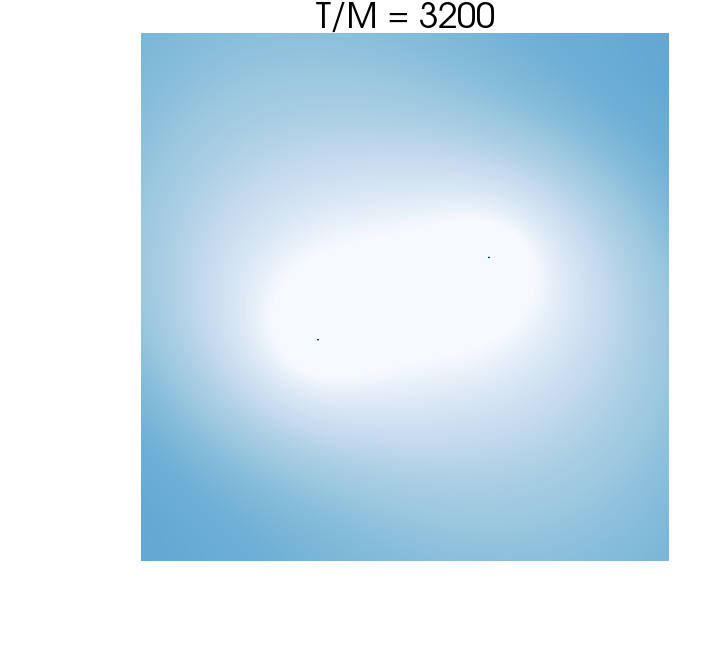}\\
\includegraphics[clip, width=3.5cm]{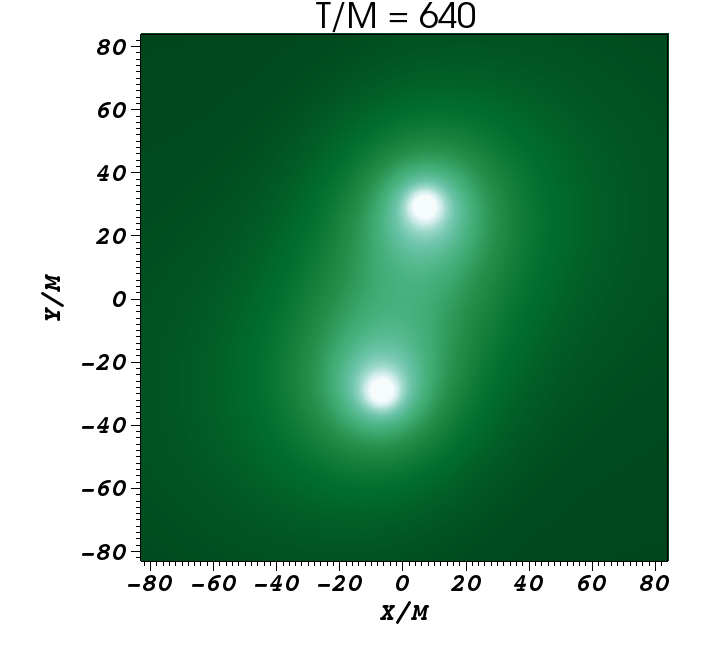}
\includegraphics[clip, width=3.5cm]{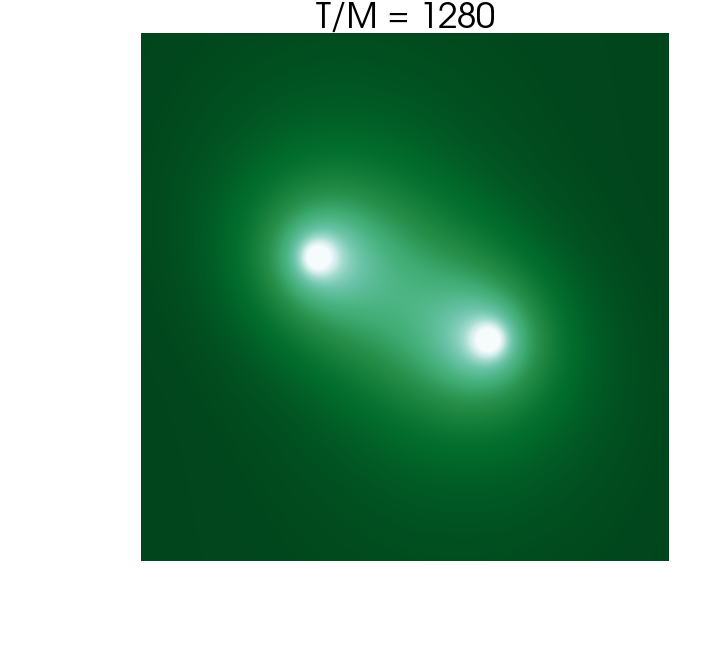}
\includegraphics[clip, width=3.5cm]{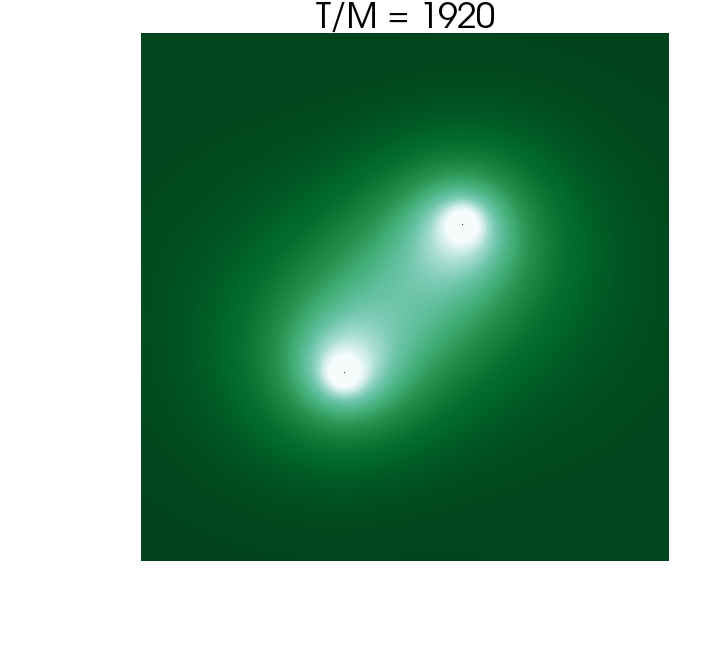}
\includegraphics[clip, width=3.5cm]{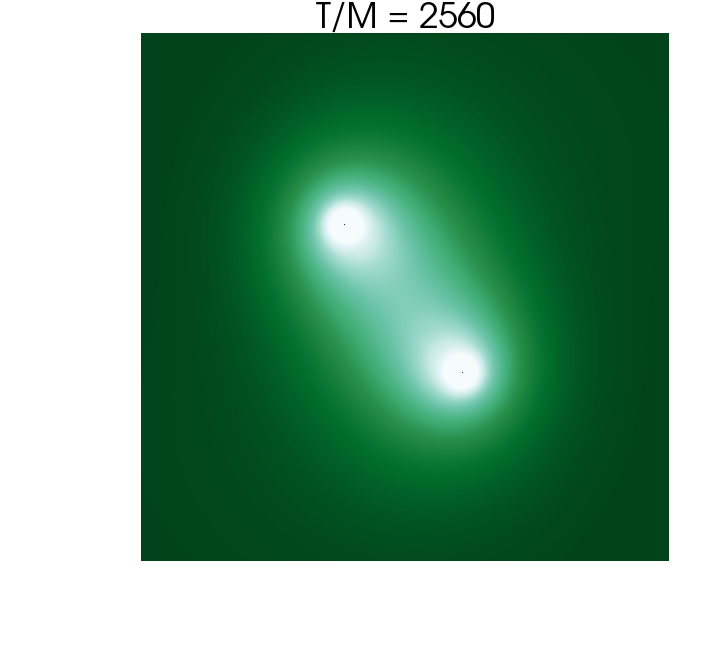}
\includegraphics[clip, width=3.5cm]{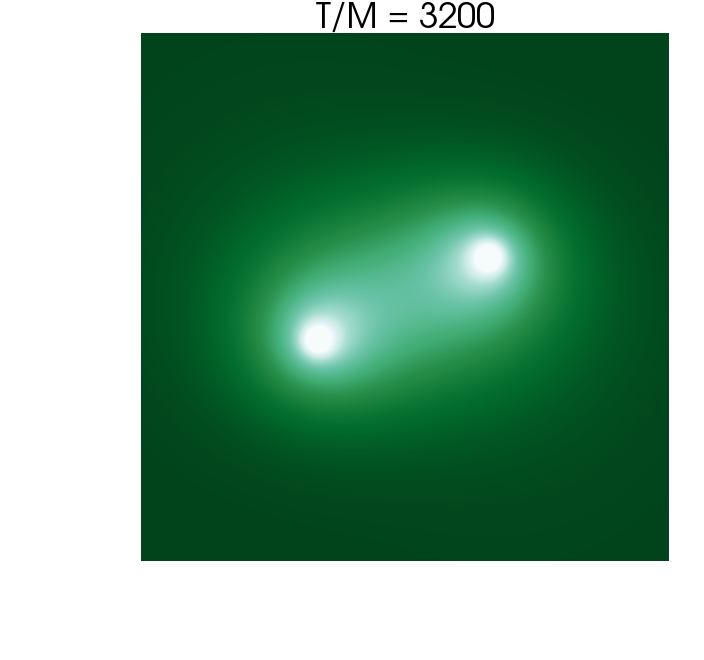}
\caption{Snapshots depicting the evolution of a scalar field around an equal-mass BHB.
  The first two rows correspond to evolutions of configuration \texttt{nonspin2} (cf.\ Table~\ref{table:simulations1}), whereas the last two rows depict evolutions of configuration \texttt{nonspin3}.
  First and third rows: scalar field. Second and fourth rows: energy density.
At late times, the scalar and energy density profile rotates counterclockwise at a frequency equal to the binary Keplerian orbital frequency.
\label{snapshot_l1m1_M05_M05_D10_mu02_gaussian1_sigma25_Ac00_1_resolution1_SBP}}
\end{figure*}

The evolution of these configurations is shown in
Fig.~\ref{snapshot_l1m1_M05_M05_D10_mu02_gaussian1_sigma25_Ac00_1_resolution1_SBP}.
Perhaps the most evident aspect of these simulations is that there is a persistent structure, a cloud or quasibound state of scalar field around the binary for 
relatively long time. In the context of Fig.~\ref{snapshot_l1m1_M05_M05_D10_mu02_gaussian1_sigma25_Ac00_1_resolution1_SBP}, the evolution timescale is large enough that the binary performed over ten periods for  configuration \texttt{nonspin2} and two periods for configuration \texttt{nonspin3}. This timescale is orders of magnitude larger than the free fall time, and yet the scalar structure persists throughout the evolution.

\begin{figure*}[tbp]
\includegraphics[width=0.45\textwidth]{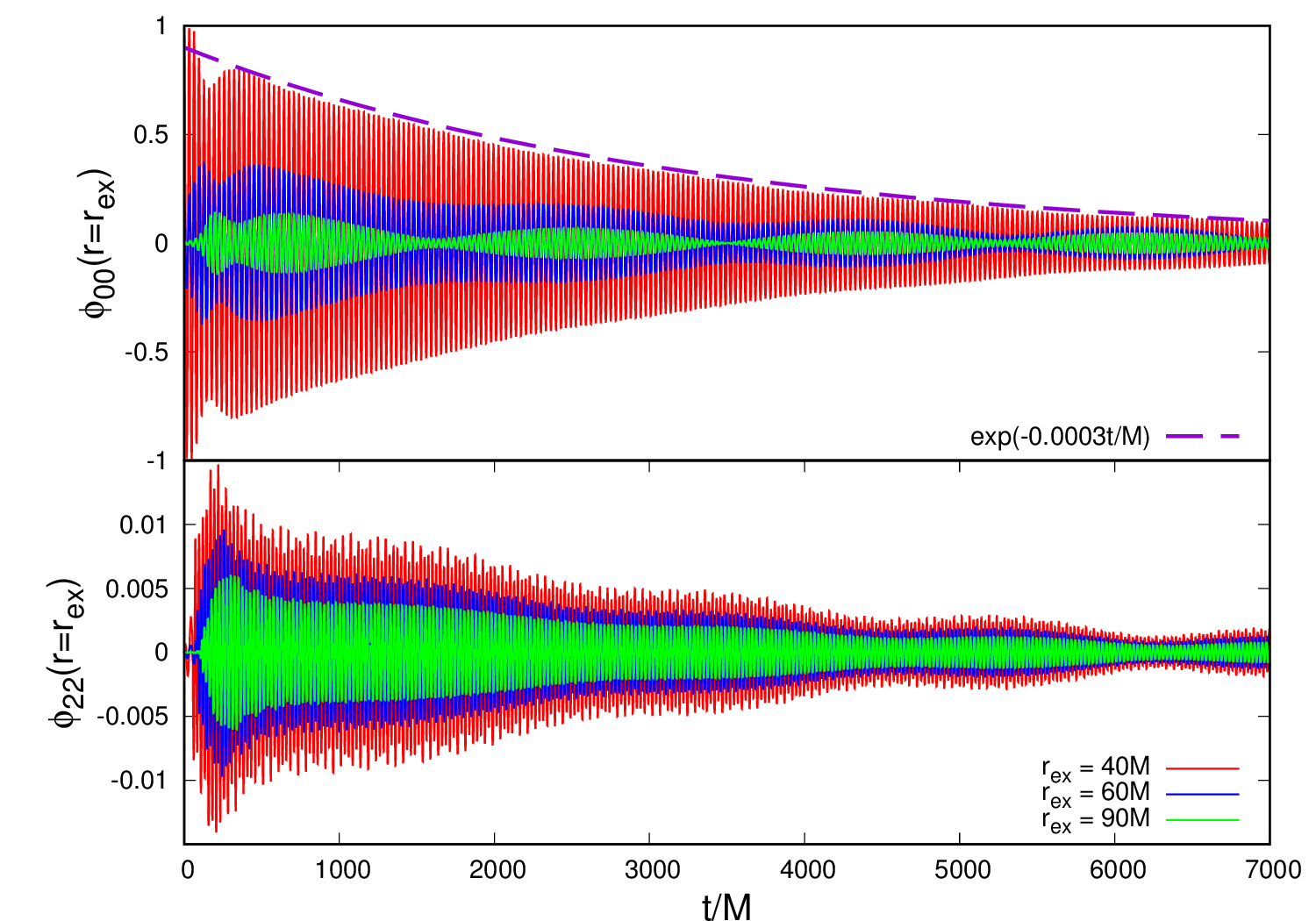} \quad
\includegraphics[width=0.45\textwidth]{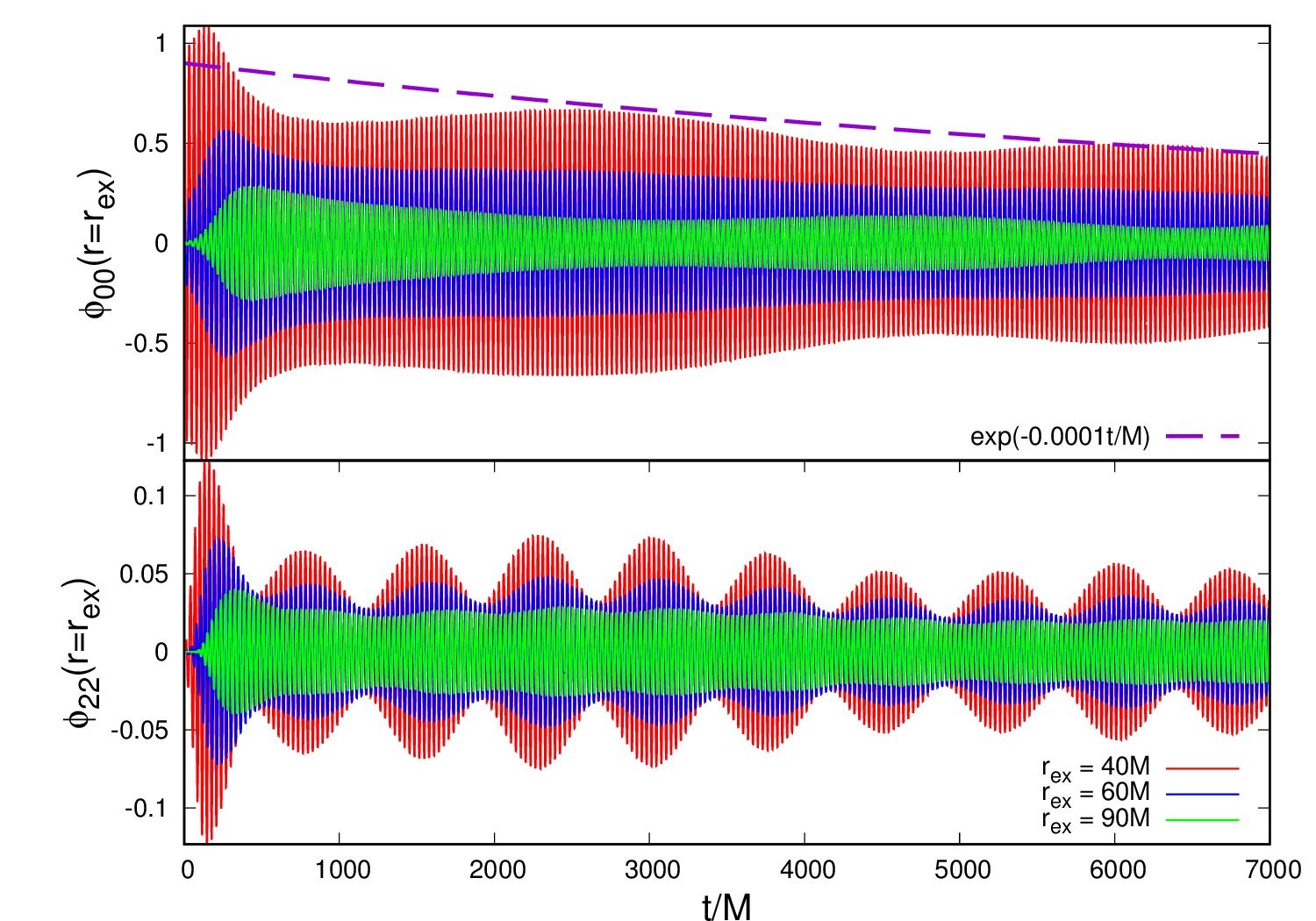}
\caption{Monopolar $l=m=0$ and dipolar $l=m=2$ components of the scalar, at selected extraction radius $r_{\rm ex}$ are shown for configuration \texttt{nonspin2} (left panel)
and configuration \texttt{nonspin3} (right panel) of Table~\ref{table:simulations1}.
The signal modulation is not due to beating of higher overtones, but to the binary orbital motion. The modulation frequency is $2m\Omega$ to a good approximation.
\label{mp_Phi_time_evolution_multi_l0m0_l2m2_M05_M05_D10_mu02_gaussian1_sigma25_Ac00_1_resolution1_SBP}}
\end{figure*}

The second noteworthy aspect is that the state keeps, roughly, the symmetry of the initial conditions. This is clearly seen in the energy density plots in Fig.~\ref{snapshot_l1m1_M05_M05_D10_mu02_gaussian1_sigma25_Ac00_1_resolution1_SBP}.
However, it is clear also that fine structure arises, clearly seen for larger binary separations, excited by the presence of the binary, an asymmetric perturber.
In particular, we observe the excitation of the quadrupole mode by the BHB. This is clearly shown in
Fig.~\ref{mp_Phi_time_evolution_multi_l0m0_l2m2_M05_M05_D10_mu02_gaussian1_sigma25_Ac00_1_resolution1_SBP},
where we show the monopole $l=m=0$ and quadrupole $l=m=2$ components of the field at selected ``extraction'' $r_{\rm ex}/M=40,\,60,\,90$.

Figure~\ref{mp_Phi_time_evolution_multi_l0m0_l2m2_M05_M05_D10_mu02_gaussian1_sigma25_Ac00_1_resolution1_SBP} shows that
this is indeed a quasibound state, decaying exponentially in time, albeit on long timescales. This is exactly as expected from
an analysis of massive fields around nonspinning BHs~\cite{Brito:2015oca}. The scalar at late times behaves as an exponentially damped sinusoid, as expected for 
quasibound states. A fit to the late-time behavior (see Fig.~\ref{mp_Phi_time_evolution_multi_l0m0_l2m2_M05_M05_D10_mu02_gaussian1_sigma25_Ac00_1_resolution1_SBP}) shows that the lifetime of such structures is $3\times 10^{3}M$ for $D=10M$, and $10^{4}M$ for $D=60M$. 
These scales should be compared with the analytical prediction, Eq.~(4.13) in Ref.~\cite{Wong:2020qom}. That expression is applicable to $D=10M$ and yields a timescale $\sim 3.9\times 10^3M$, in excellent agreement with our numerical results.

\begin{table}[!htb]
\caption{Spectrum content of waveforms and comparison against nonrelativistic results. The third column of this table shows the location of the dominant peak of the waveform, from time evolutions, in
Fourier space (for $l=m=1,\,2$ there are two dominant peaks). 
The fourth column shows the nonrelativistic prediction, obtained by solving the coupled system Eq.~\eqref{eq:spheroidals} (these values are also in Table~\ref{table:eigen_spheroidal} in a slightly different form), which is formally equivalent to solving the dihydrogen molecule. As we noted before, the fundamental mode $E_{000}$ should always be present in each spherical harmonic $(l,\,m)$ basis used for the simulations. Other components are also present, but we find those to be subdominant. 
The agreement between both is very good and lends strong support to the interpretation that these are ``molecular'' gravitational quasibound states.
\label{table:FT}
}

\begin{ruledtabular}
\begin{tabular}{cccc}
% \hline
Run        &$(l,\,m)$     &$M\omega $      & $M\left(\mu+E_{000}\pm m\Omega\right)$\\
\hline
\texttt{nonspin3} & $(0,\,0)$ &0.1976          & 0.1973\\
\hline
\texttt{spin2} & $(1,\,1)$ &0.1992          & 0.1994\\
%\hline
               &      &0.1948          & 0.1951\\
\hline
\texttt{nonspin3} & $(2,\,2)$ &0.2012          & 0.2016\\
%\hline
   & &0.1930          & 0.1930\\
% \hline
\end{tabular}
\end{ruledtabular}
\end{table}
In accordance with our analysis in Sec.~\ref{section:non_relativistic}, the scalar field is oscillating with a frequency $\sim \mu$. To quantify the agreement with the nonrelativistic analysis, we computed the Fourier spectrum of the monopole and quadrupole modes at different radii, and averaged the result. We find clear peaks at different frequencies, the dominant ones are shown in
Table~\ref{table:FT} for $D=60M$ and compared against the nonrelativistic prediction of Sec.~\ref{section:non_relativistic}.\footnote{One needs to pay attention when comparing against the results of Sec.~\ref{section:non_relativistic}, since a rotation of axes is involved and, as we mentioned, at finite separations
a spherical harmonic mode maps into a superposition of the $(m_{\xi},\,m_\eta,\,m_\chi)$ states used in Table~\ref{table:eigen_spheroidal}.} The agreement is of order 0.1\% or better, providing strong evidence that
bound, molecular-like gravitational states do form around BHBs.
Notice also that we find two (or more) peaks for $l=m$ modes. For $D=60M$ we can read from the table that their separation in frequency $\Delta^\omega_{m}\equiv M(\omega_{m}-\omega_{-m})$ is $\Delta^\omega_{m=2}=0.0082$, in good agreement with the expected prediction of Sec.~\ref{subsec:rotating_frame_wave}, $\Delta^\omega_{m}=2mM\Omega$ ($M\Omega\sim 0.0022$ for this example).

\begin{figure*}[tbp]
\includegraphics[width=0.45\textwidth]{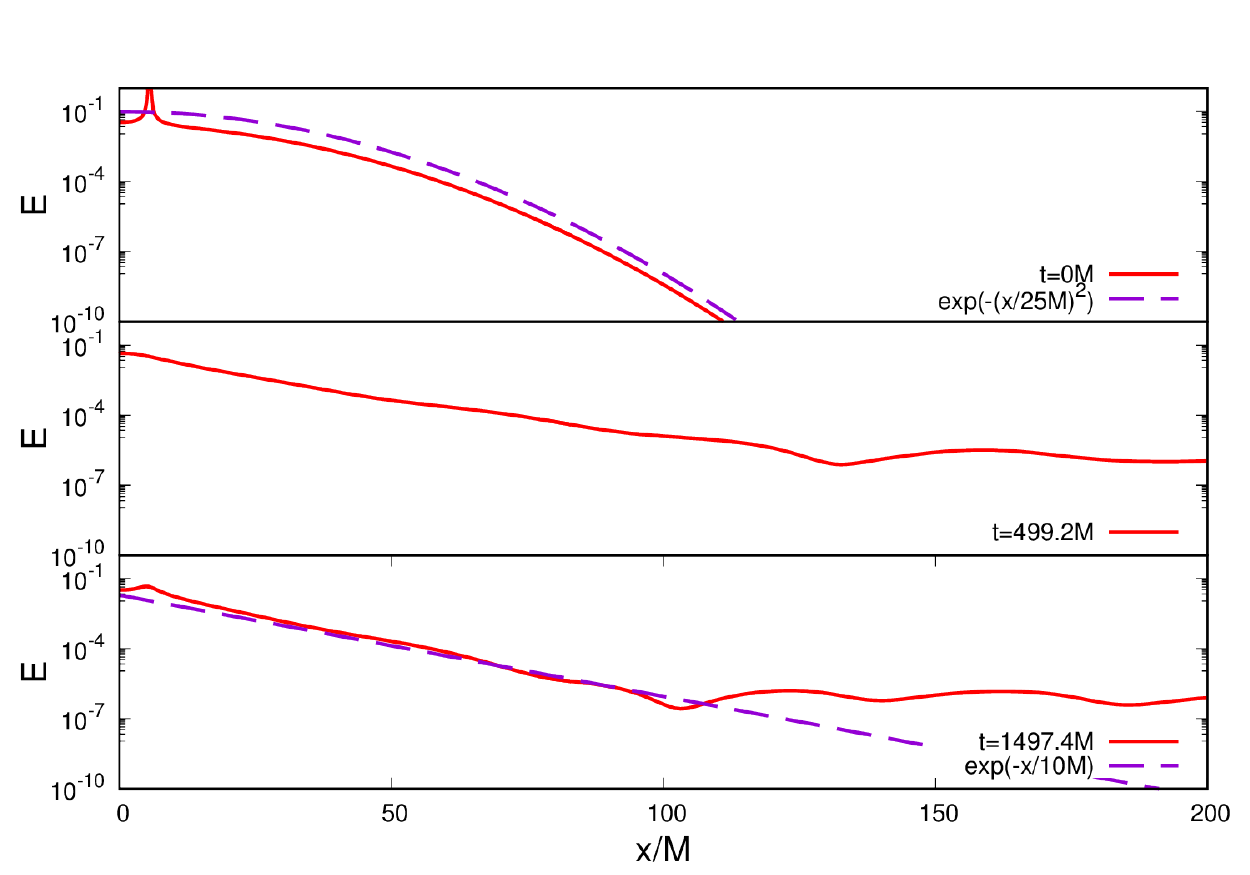} \quad
\includegraphics[width=0.45\textwidth]{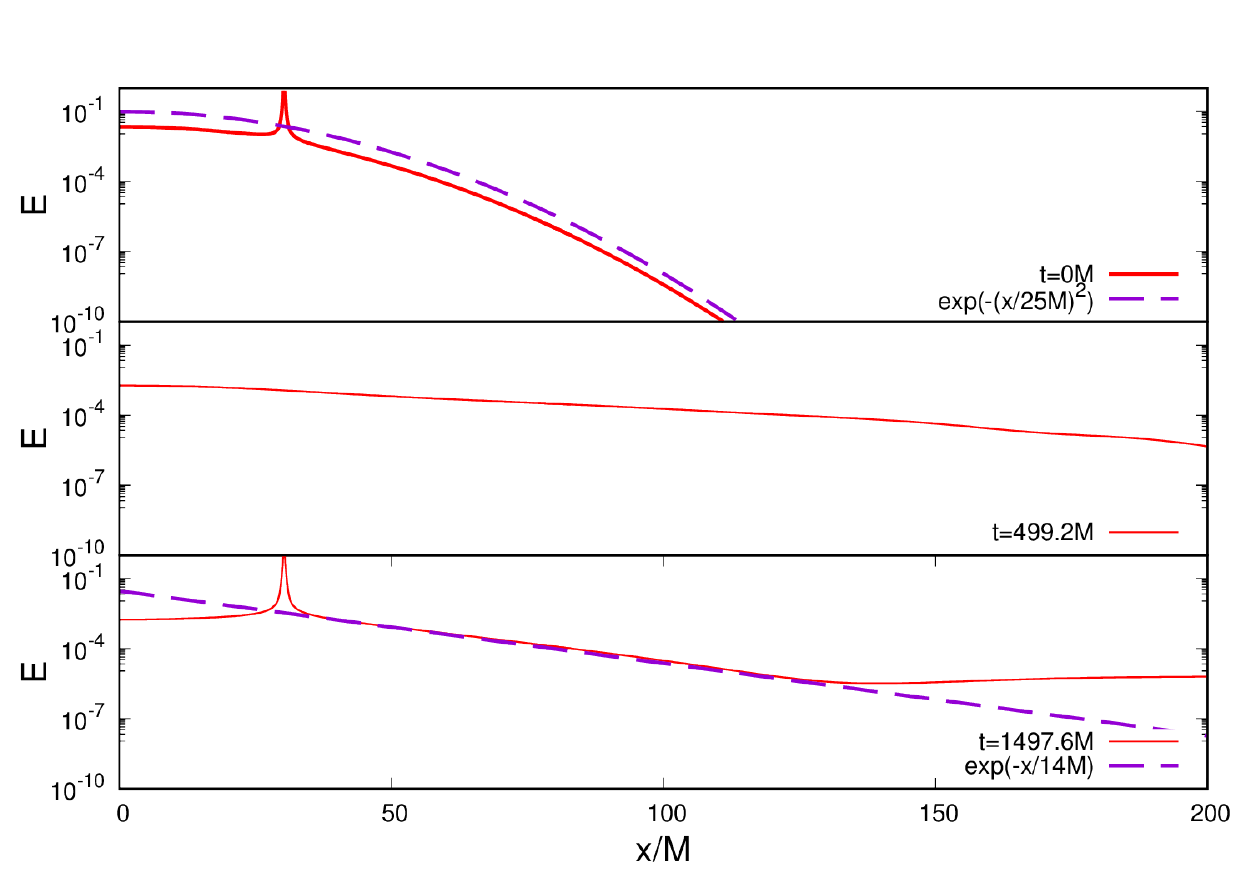}
\caption{Energy density of the scalar field on the $x$ axis for the configurations \texttt{nonspin2} (left panel) and \texttt{nonspin3} (right panel). Dashed lines are best fits to the numerical results, and agree well with the analytical, nonrelativistic predictions (see text).
\label{time_evolution_energydensity_M05_M05_D10_mu02_gaussian1_sigma25_Ac00_1_resolution1_SBP_v2}}
\end{figure*}

The interpretation of these states as molecular-like is further supported by the spatial profile of the scalar and energy density, shown in
Fig.~\ref{time_evolution_energydensity_M05_M05_D10_mu02_gaussian1_sigma25_Ac00_1_resolution1_SBP_v2}. The time evolution of the energy density along the $x$ axis is depicted in the three panels of this figure.
The late-time profile around the binary is well described by a density $\sim e^{-2r/(25M)}$. This profile is in accord with the nonrelativistic, single BH expression [Eq.~\eqref{eq:atom_limit_wave}]
for the quasibound state. Our results indicate that this is also a good expression, even for the moderately large separations that we studied.
Figure~\ref{time_evolution_energydensity_M05_M05_D10_mu02_gaussian1_sigma25_Ac00_1_resolution1_SBP_v2} shows that the exponential falloff gives rise to a slowly decaying but small tail of energy density for distances $r\gtrsim 100 M$. This looks to be a slow, radiative component part of the signal.

In fact, this gravitational system behaves like a rotating molecule in quantum mechanics.\footnote{Disclaimer: one should not read too much in this analogy. The shared electron cloud in a molecule is what binds the atoms together. Here, the ``atoms'' (two BHs) are bound together by gravity and then the ``electron cloud'' is a solution of the scalar wave equation on that background.}  The snapshots of the scalar field and energy density shown in Fig.~\ref{snapshot_l1m1_M05_M05_D10_mu02_gaussian1_sigma25_Ac00_1_resolution1_SBP} show that the scalar field is not only localized around the BHB, but that it is dragged by it, rotating counterclockwise, along the orbital motion of the binary. The field shows modulations at low-frequencies---in particular, at $2m\Omega$---as can be seen in the figures. Such modulation is the expected for an equal-mass binary. Notice that the signal is almost equally modulated in amplitude at different extraction radii; thus the low-frequency envelope is {\it not} the result of beatings caused by overtone excitation. In other words the scalar is indeed gravitationally dragged by the binary. The period of such a pattern is the same as the orbital period of the binary.

%%%%%%%%%%%%%%%%%%%%%%%%%%%%%%%%%%%%%%%%%%%%%%%%%%%%%%%%%%%%%%%%%%%%%%%%%%%
\subsection{Corotating dipolar global states}
%%%%%%%%%%%%%%%%%%%%%%%%%%%%%%%%%%%%%%%%%%%%%%%%%%%%%%%%%%%%%%%%%%%%%%%%%%%
%
\begin{figure}[htb]
\includegraphics[clip, width=4.0cm]{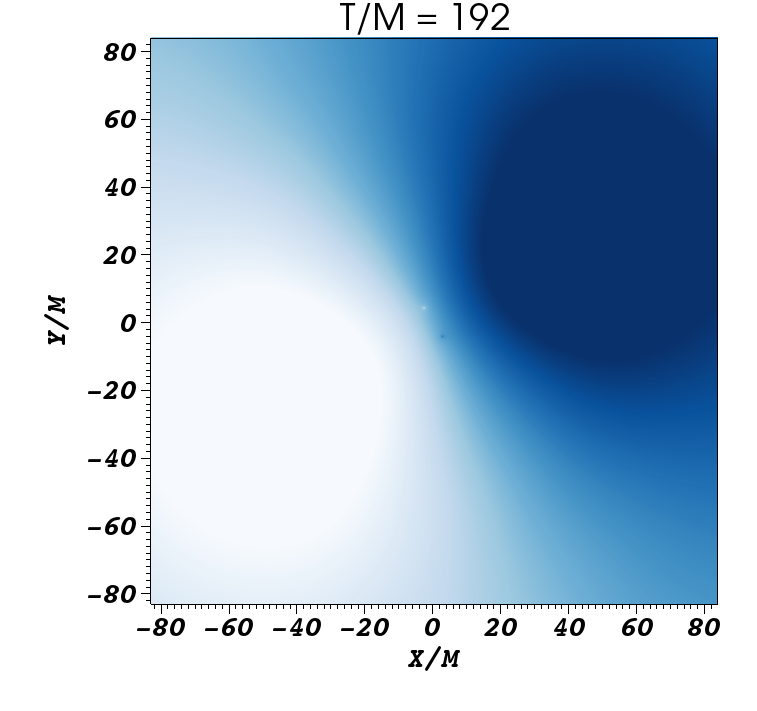}
\includegraphics[clip, width=4.0cm]{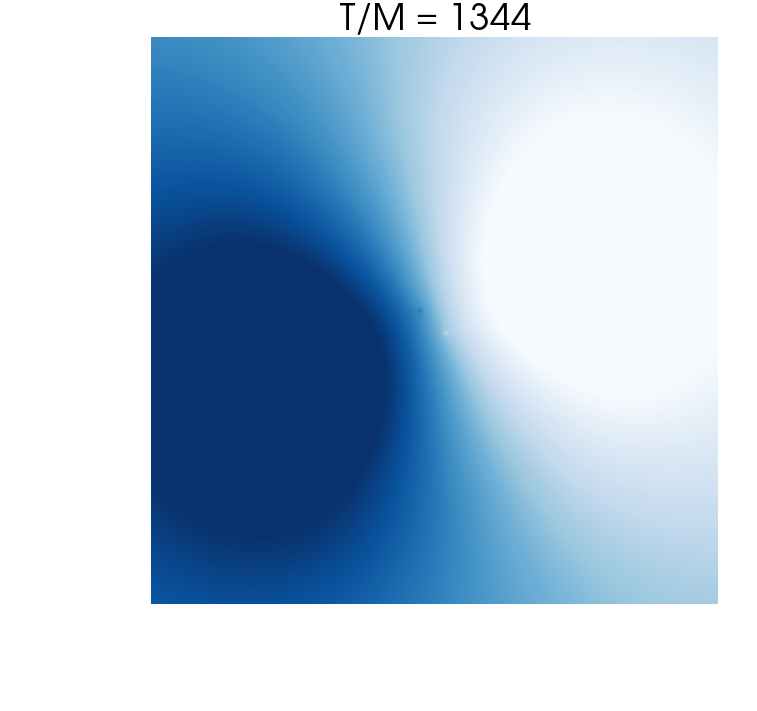}
\\
\includegraphics[clip, width=4.0cm]{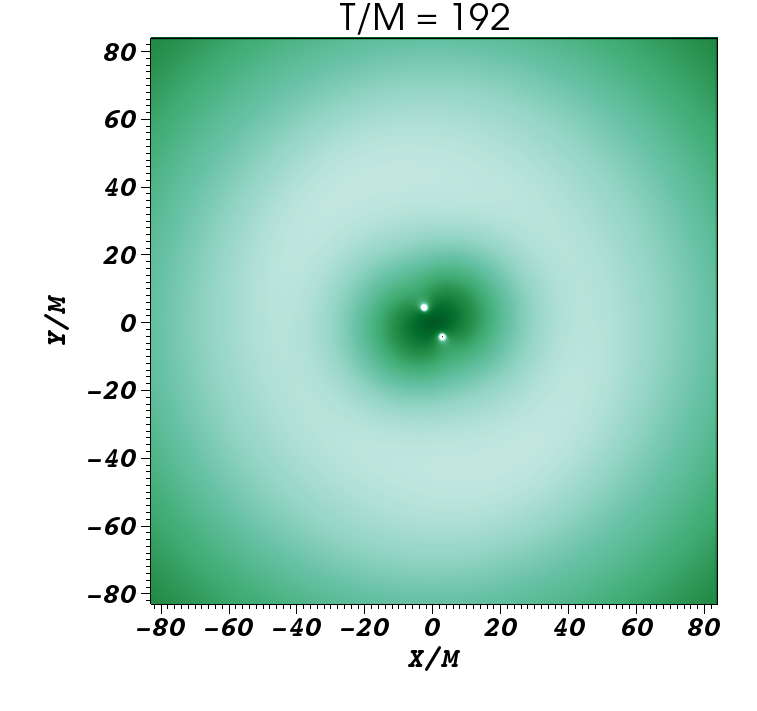}
\includegraphics[clip, width=4.0cm]{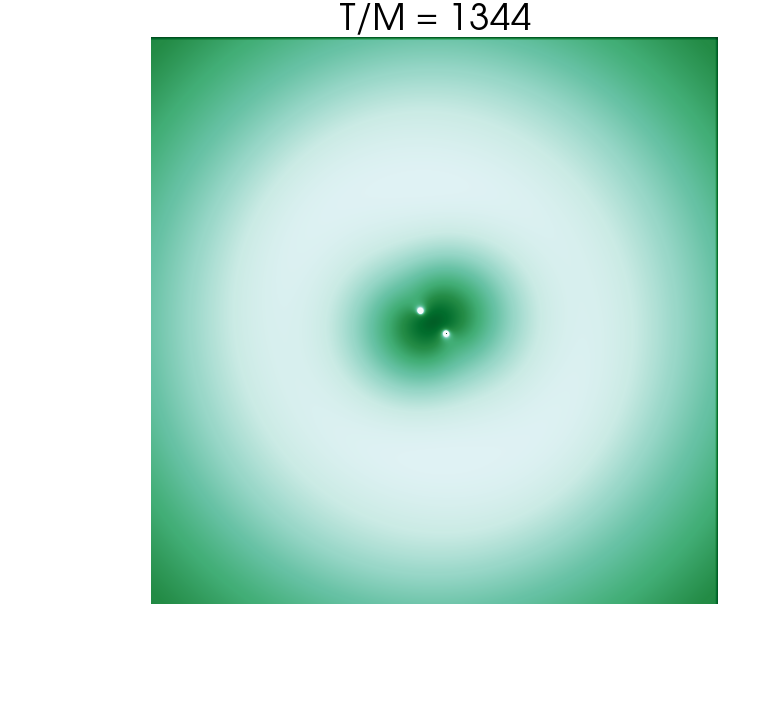}
\caption{Snapshots depicting the evolution of a scalar around an equal-mass BHB for configuration \texttt{spin1} of Table~\ref{table:simulations2}.
The initial conditions are therefore those of a dipolar $l=1,\,m=-1$ scalar configuration corotating with the binary in a counterclockwise direction.
Top panels: evolution of scalar field.
Bottom panels: evolution of energy density.
At late times, the scalar and energy density profile rotate counterclockwise at a frequency dictated by the binary orbital frequency.
\label{snapshot_l1m1_M05_M05_D10_mu02_clump_r_exp_omega020_width25_radius0_l1m1_resolustion1_SBP}}
\end{figure}
\begin{figure}[htb]
\includegraphics[clip, width=4.0cm]{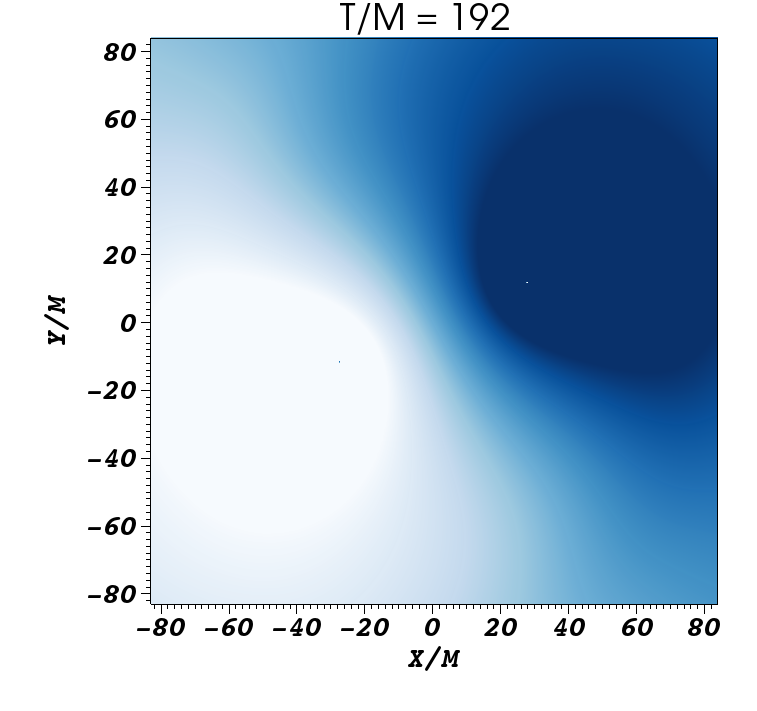}
\includegraphics[clip, width=4.0cm]{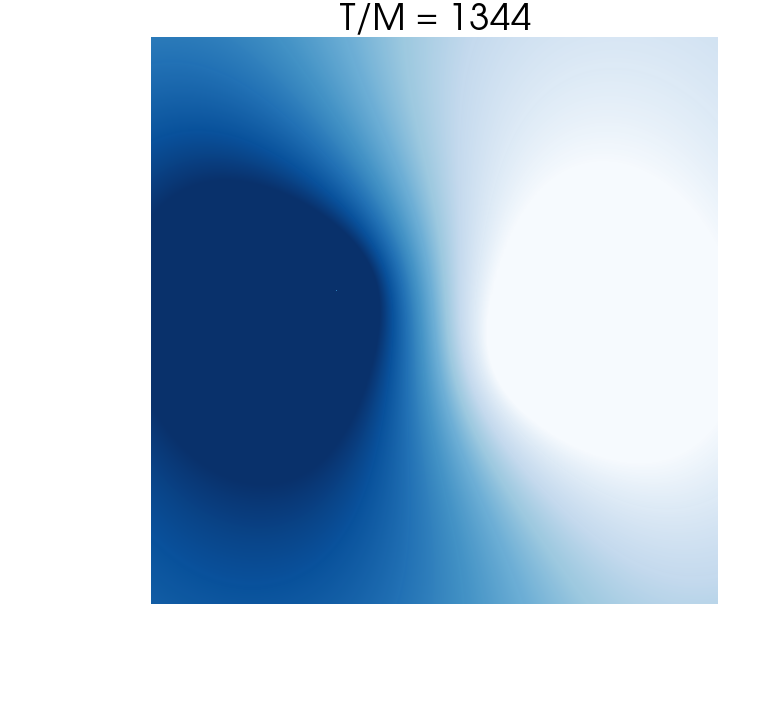}
\\
\includegraphics[clip, width=4.0cm]{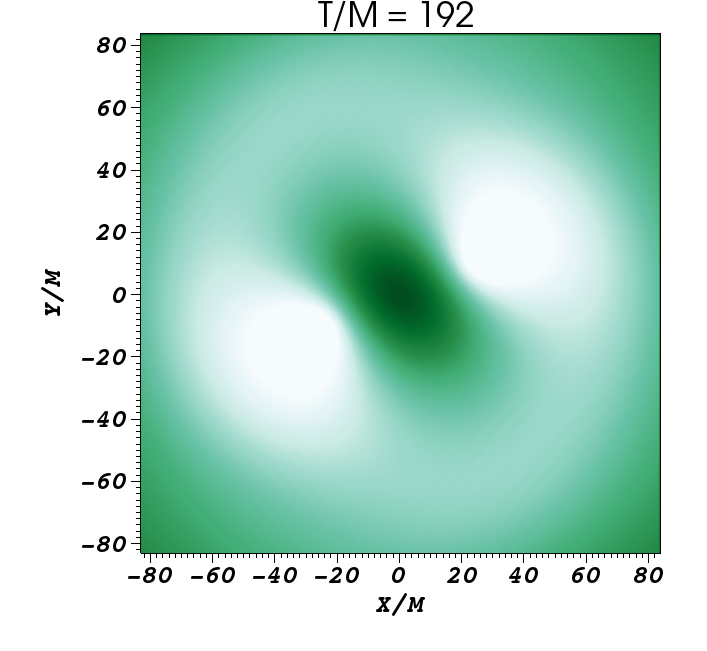}
\includegraphics[clip, width=4.0cm]{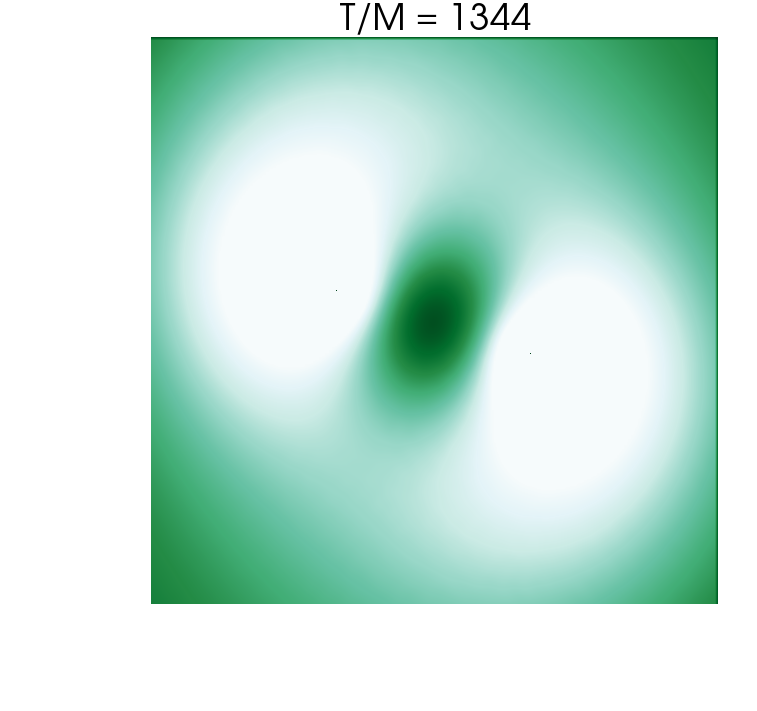}
\caption{Same as Fig.~\ref{snapshot_l1m1_M05_M05_D10_mu02_clump_r_exp_omega020_width25_radius0_l1m1_resolustion1_SBP}, but for configuration \texttt{spin2} of Table~\ref{table:simulations2}.
\label{snapshot_l1m1_M05_M05_D60_mu02_clump_r_exp_omega020_width25_radius0_l1m1_resolustion1_SBP}}
\end{figure}
The previous evolutions referred to spherically symmetric, momentarily static initial data. We now consider time-asymmetric initial data. 
Let us focus on dipolar initial data corotating with the binary---i.e.,
configurations \texttt{spin1} and \texttt{spin2} of Table~\ref{table:simulations2}.
The evolution of the spatial profile of the field and its energy density is shown in Figs.~\ref{snapshot_l1m1_M05_M05_D10_mu02_clump_r_exp_omega020_width25_radius0_l1m1_resolustion1_SBP} and \ref{snapshot_l1m1_M05_M05_D60_mu02_clump_r_exp_omega020_width25_radius0_l1m1_resolustion1_SBP}. A dipolar pattern stands out from these plots.

Figures~\ref{snapshot_l1m1_M05_M05_D10_mu02_clump_r_exp_omega020_width25_radius0_l1m1_resolustion1_SBP} and \ref{snapshot_l1m1_M05_M05_D60_mu02_clump_r_exp_omega020_width25_radius0_l1m1_resolustion1_SBP} show that for compact binaries (when the BH separation is smaller than the typical size of the cloud),
the energy density of the state acquires a torus-like shape, centered on the binary, and supported by its angular momentum. This is similar to the topology of quasibound states around single BHs.
On the other hand, for large BH separations as in Fig.~\ref{snapshot_l1m1_M05_M05_D60_mu02_clump_r_exp_omega020_width25_radius0_l1m1_resolustion1_SBP}, the profile is no longer connected; the torus ``breaks up'', and leaves two overdensity clumps of scalar field corotating with the BHB.

\begin{figure}[th]
\includegraphics[width=0.45\textwidth]{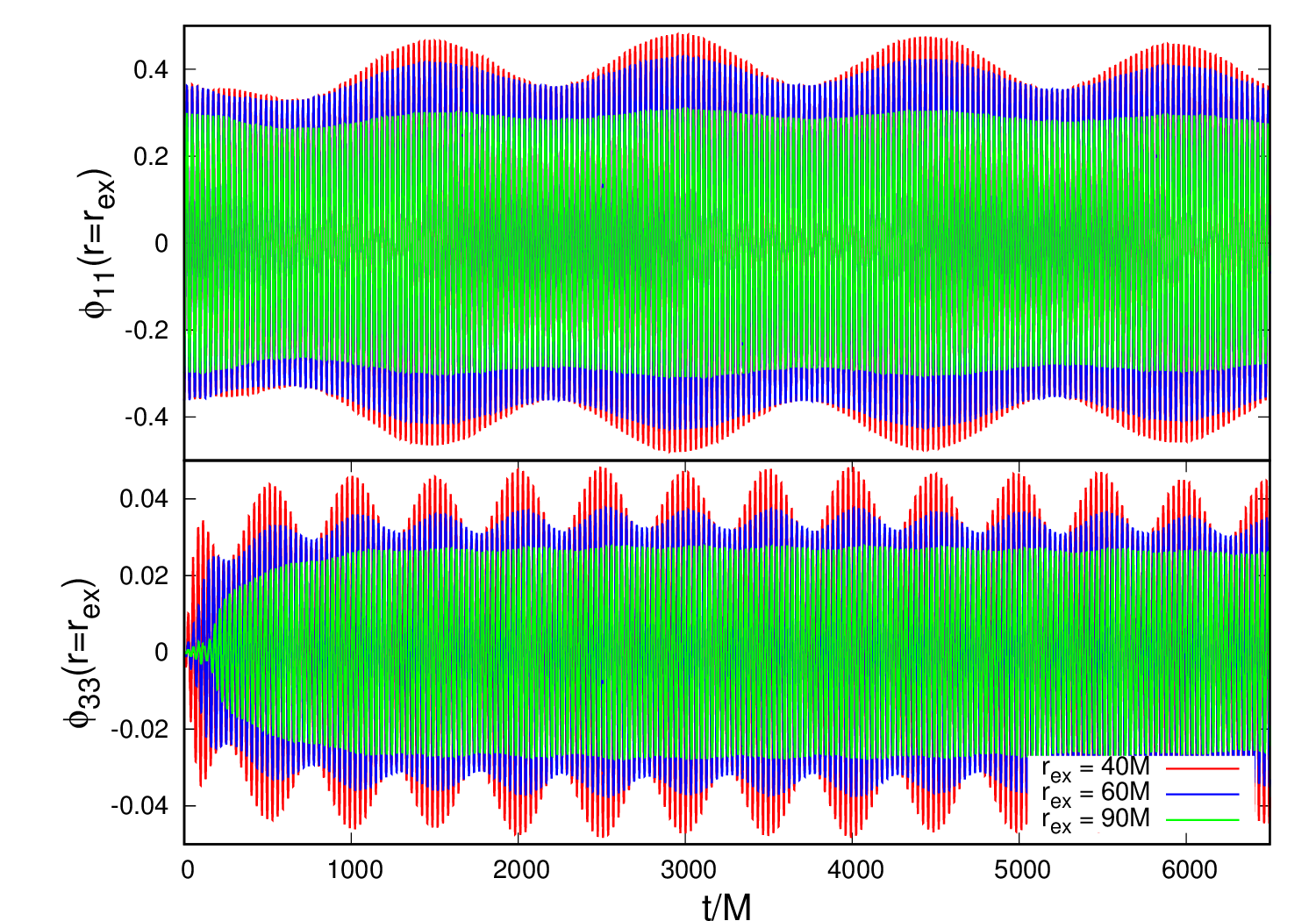}
\caption{
 Dipolar ($l=m=1$) and octupolar ($l=m=3$) components of the scalar field, measured at various $r_{\rm ex}$, for configuration \texttt{spin2} of Table~\ref{table:simulations2}.
 Notice that a dipolar mode is already present initially.
 This is the same initial data as that of  Fig.~\ref{snapshot_l1m1_M05_M05_D60_mu02_clump_r_exp_omega020_width25_radius0_l1m1_resolustion1_SBP}.
 The signal modulation is not due to the beating of higher overtones, but due to the binary orbital motion. The modulation frequency is $2m\Omega$ to a good approximation.
\label{mp_Phi_time_evolution_multi_l1m1_l3m3_M05_M05_D60_mu02_clump_r_exp_omega020_width25_radius0_l1m1_resolustion1_SBP.pdf}
}
\end{figure}
The presence of the binary excites other modes with similar symmetries to that in the initial data. In particular, we see a strong octupole $l=m=3$ mode, shown in Fig.~\ref{mp_Phi_time_evolution_multi_l1m1_l3m3_M05_M05_D60_mu02_clump_r_exp_omega020_width25_radius0_l1m1_resolustion1_SBP.pdf} for $D=60 M$.
Notice that the octupolar mode grows from a negligible value to roughly 10\% in amplitude of the dipolar component.
Notice also the large timescales involved: the amplitude of these components is approximately constant up to timescales of order $\sim 10^4M$ or larger.
The modulation in the signal is, as we explained before, due to the motion of the binary, and has a frequency $2m\Omega$ as expected for an equal-mass binary.

These results indicate that the evolution drove the system to a quasibound state, a relativistic analogue of the molecular solutions discussed in Sec.~\ref{sec:equivalence_molecule} for the nonrelativistic system. Together with the previous results, and as we will insist below, these features indicate that the formation of quasibound states is a robust result for general initial conditions.
The analysis of the Fourier-decomposed signal shows a frequency content which is peaked at $M\omega=0.1992,\,0.1948$. This is in good agreement with the nonrelativistic predictions of Sec.~\ref{sec:equivalence_molecule} (which yield $0.1994$ and $0.1951$, respectively). One finds (see Table~\ref{table:FT}) $\Delta^\omega_{m=1}=0.0044$, also in very good agreement with the expectations.
\begin{figure}[th]
\includegraphics[width=0.45\textwidth]{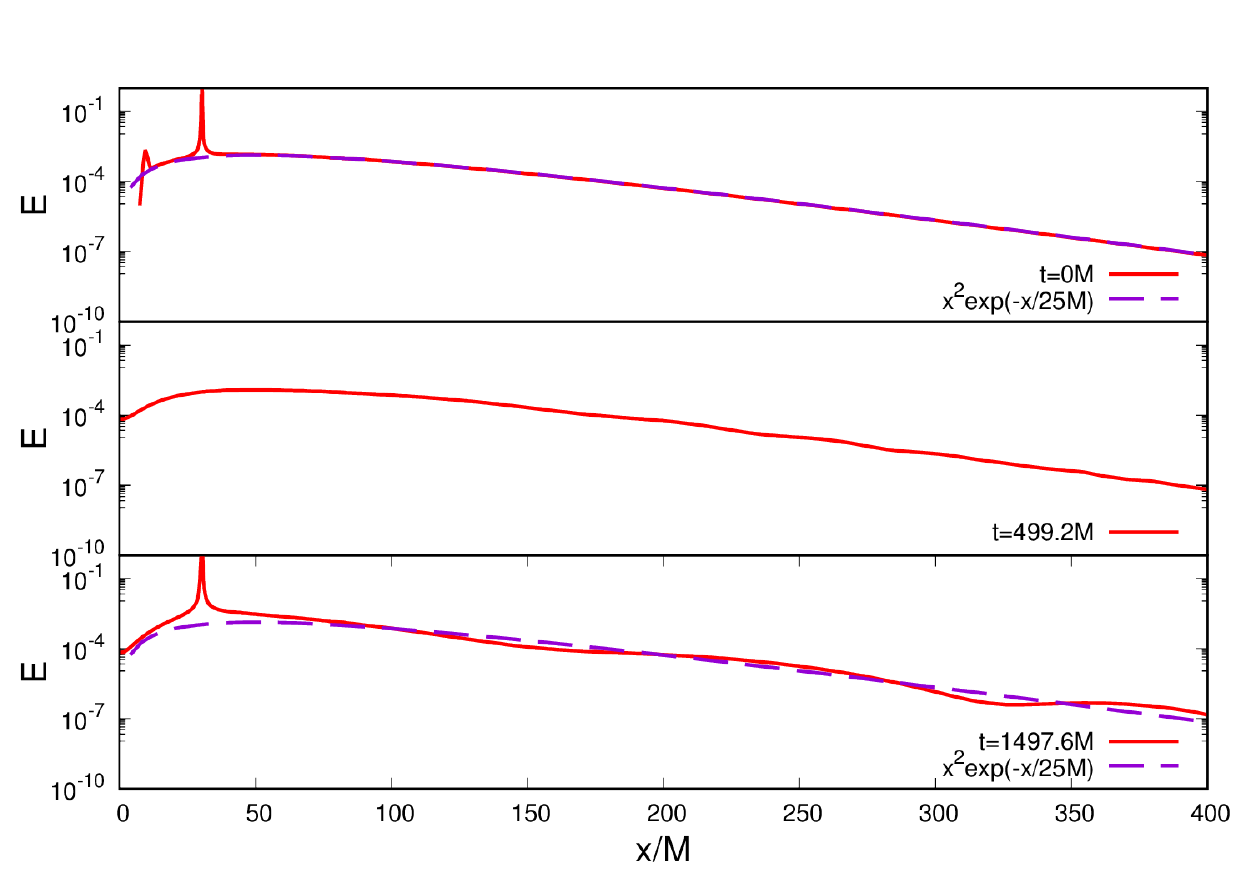}
\caption{Energy density of a quasibound state around a BHB separated by $D=60M$. The density is measured along the $x$ axis. Initial data is that of configuration \texttt{spin2} in Table~\ref{table:simulations2}, corresponding to a corotating dipole.
The density was measured at different instants, and our results show that the late-time profile is in accordance
with the nonrelativistic prediction for a bound state.
\label{time_evolution_energydensity_M05_M05_D10_mu02_clump_r_exp_omega020_width25_radius0_l1m1_resolustion1_SBP_v2}}
\end{figure}
The energy density of the configuration is shown in Fig.~\ref{time_evolution_energydensity_M05_M05_D10_mu02_clump_r_exp_omega020_width25_radius0_l1m1_resolustion1_SBP_v2}.
These results show clearly that the initial conditions are similar to those of a quasibound state, and the system evolution does not take it away significantly from the initial conditions.
The asymptotic behavior of the density profile is well described by $(re^{-r/50M})^{2}$. Note that the nonrelativistic prediction [Eq.~\eqref{eq:atom_limit_wave}] for the scalar profile is of the form $\sim e^{-M\mu^2r/(\ell+1)}=e^{-r/(50M)}$ for the fundamental mode when $M\mu=0.2$. This is in perfect agreement with our time-evolution results around a BHB.

%%%%%%%%%%%%%%%%%%%%%%%%%%%%%%%%%%%%%%%%%%%%%%%%%%%
\subsection{Counterrotating dipolar global states}
%%%%%%%%%%%%%%%%%%%%%%%%%%%%%%%%%%%%%%%%%%%%%%%%%%%
%
\begin{figure}[th]
\includegraphics[clip, width=4.0cm]{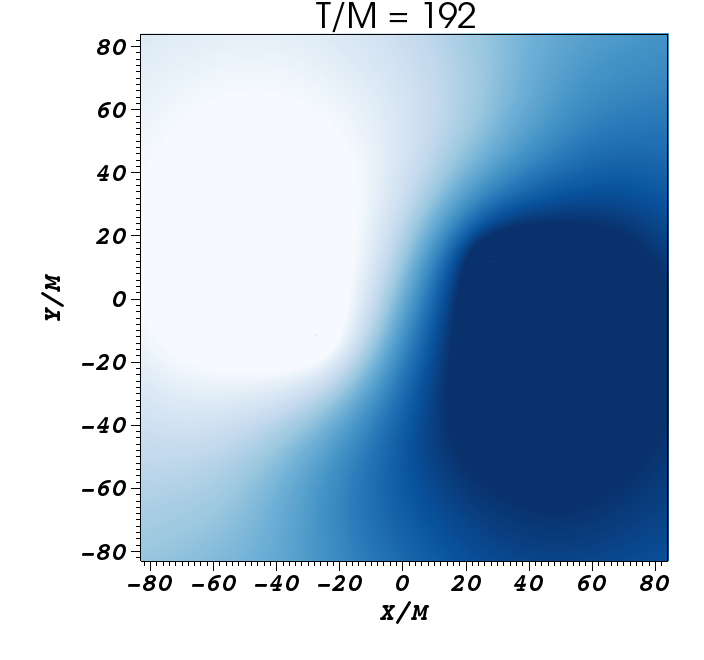}
\includegraphics[clip, width=4.0cm]{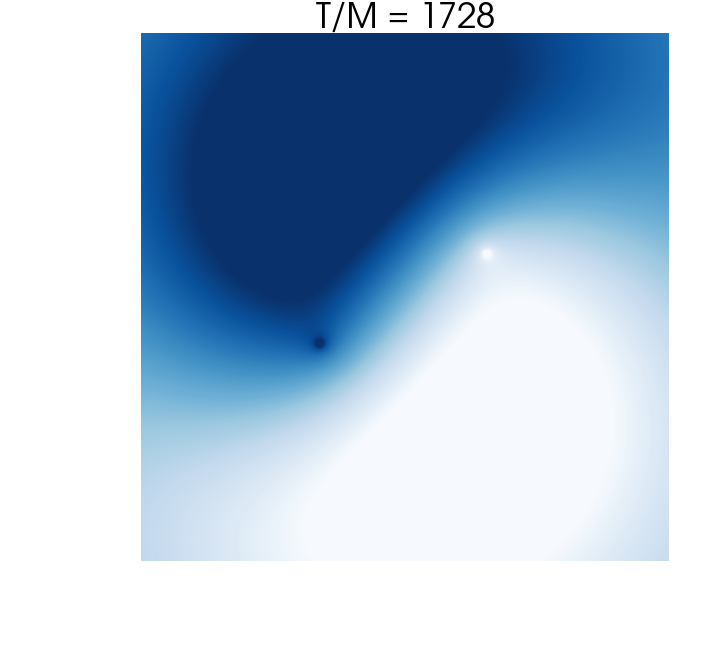}\\
\includegraphics[clip, width=4.0cm]{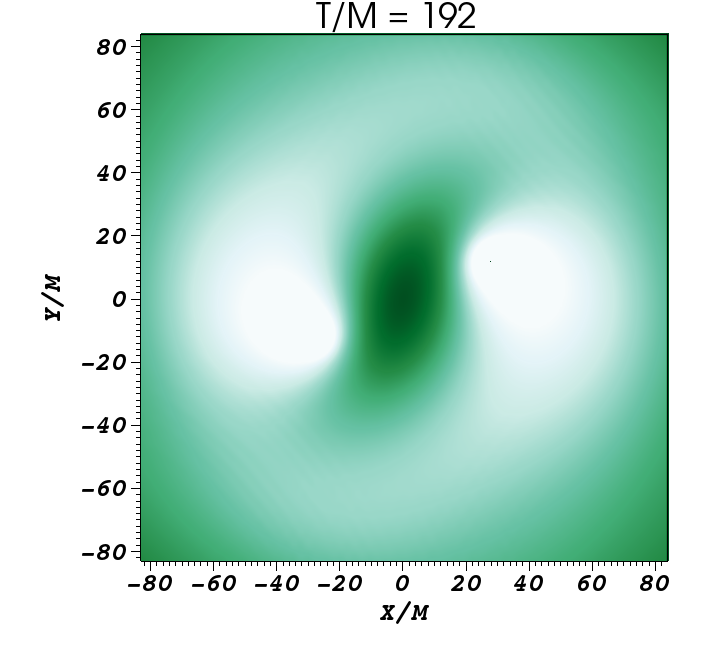}
\includegraphics[clip, width=4.0cm]{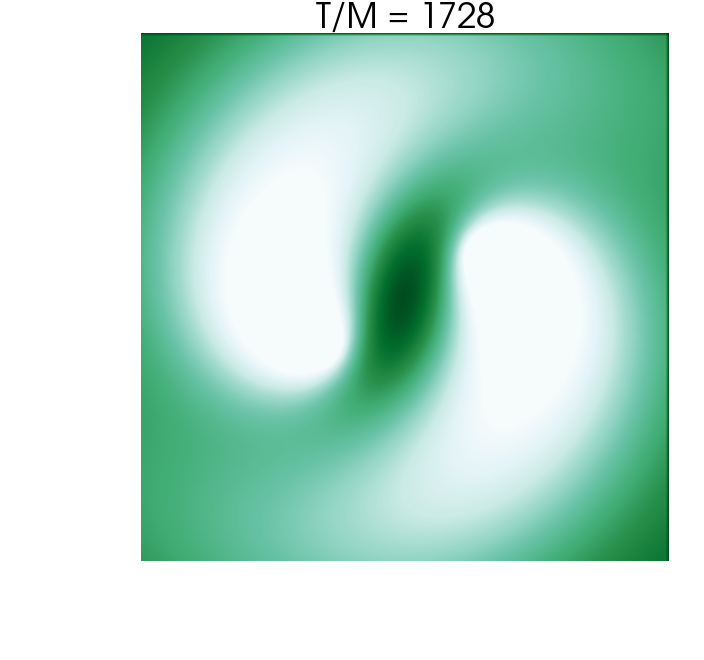}
\caption{Evolution of configuration \texttt{spin3} in Table~\ref{table:simulations2}, corresponding to a dipolar ($l=m=1$) scalar field counterrotating with the binary.
Top panels: evolution of scalar field.
Bottom panels: evolution of energy density.
\label{snapshot_l1m1_M05_M05_D60_mu02_clump_r_exp_omega020_width25_radius0_l1p1_resolustion1_SBP}}
\end{figure}
Similar results hold for the evolution of counterrotating initial data (with respect to the binary, thus data rotating clockwise).
We have evolved configuration \texttt{spin3} of Table~\ref{table:simulations2} and
the corresponding profiles of the field and energy density are shown in Fig.~\ref{snapshot_l1m1_M05_M05_D60_mu02_clump_r_exp_omega020_width25_radius0_l1p1_resolustion1_SBP}.

Since the initial profile has angular momentum which is opposite to that of the
binary, the scalar field rotates in a direction opposite to the BHB. However,
the energy density of the scalar cloud does rotate in the same direction as that
of the binary. The asymptotic late-time spatial distribution of the energy
density is well described by a radial dependence $(re^{-r/50M})^{2}$, consistent
with a nonrelativistic analysis of quasibound states.

\begin{figure}[th]
\includegraphics[width=0.45\textwidth]{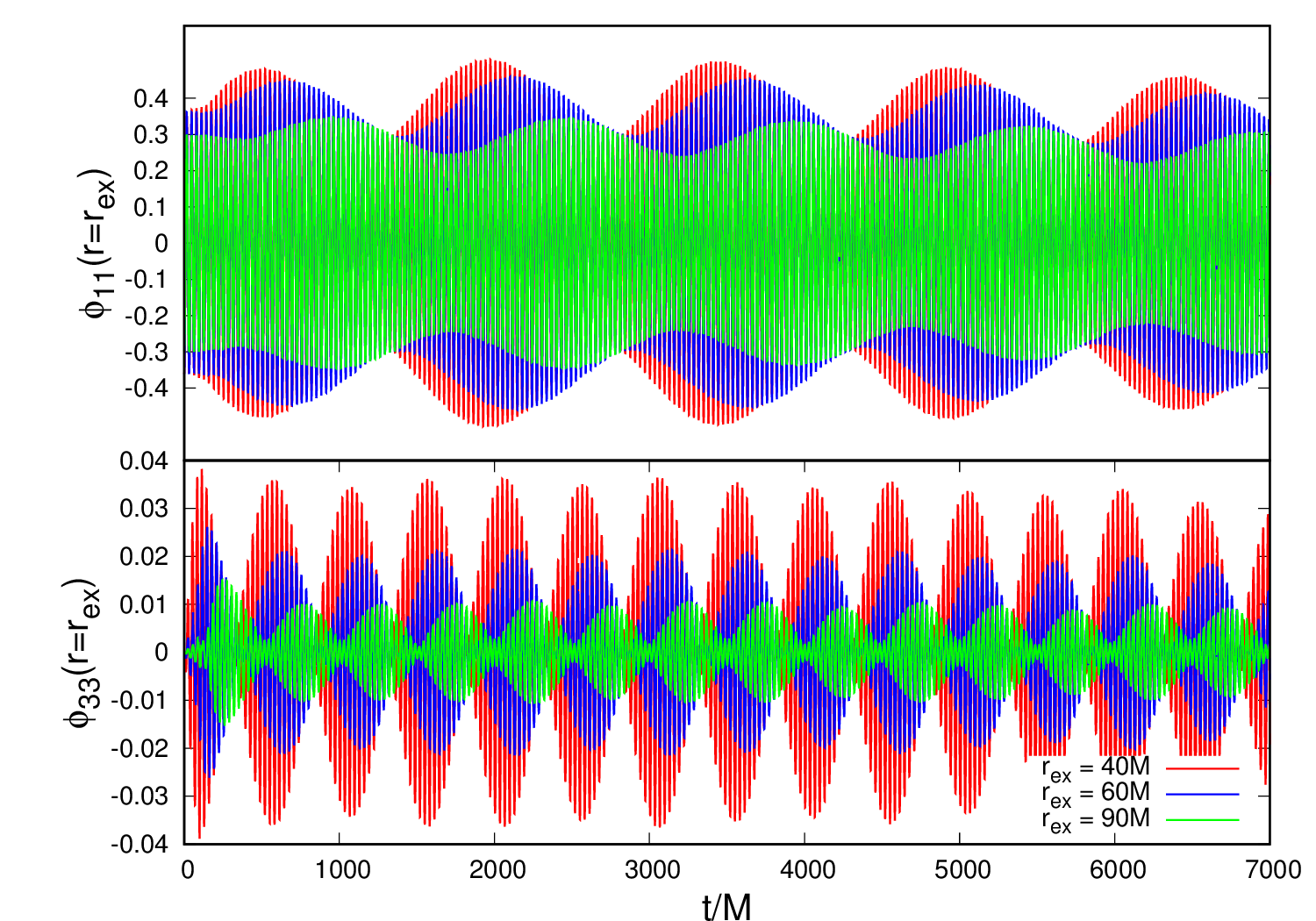}
\caption{Multipolar $l=m=1,\,3$ components of the scalar field, corresponding to configuration \texttt{spin3}. These are the same initial conditions as those of Fig.~\ref{snapshot_l1m1_M05_M05_D60_mu02_clump_r_exp_omega020_width25_radius0_l1p1_resolustion1_SBP}.
  The signal modulation is not due to the beating of higher overtones, but due to the binary orbital motion.
\label{mp_Phi_time_evolution_multi_l1m1_l3m3_M05_M05_D10_mu02_clump_r_exp_omega020_width25_radius0_l1p1_resolustion1_SBP}}
\end{figure}
As we discuss in Appendix~\ref{sec:Excited mode by binary spacetime}, selection rules imply that other modes must be excited---in particular, the $l=m=3$ mode.
Figure~\ref{mp_Phi_time_evolution_multi_l1m1_l3m3_M05_M05_D10_mu02_clump_r_exp_omega020_width25_radius0_l1p1_resolustion1_SBP} shows precisely this. 
It is also clear that a very long-lived quasibound state forms (the lifetime is so large, in fact, that we were unable to estimate it).

%%%%%%%%%%%%%%%%%%%%%%%%%%%%%%%%%%%%%%%%%%%%%
\subsection{General initial data}
%%%%%%%%%%%%%%%%%%%%%%%%%%%%%%%%%%%%%%%%%%%%%
We investigated other types of initial conditions, such as narrower pulses, with smaller width $\sigma$.
This changes the amount of field that is dissipated in the early stages, but a quasibound state always ends up forming,
with a phenomenology similar to what we described.

We also studied data with higher initial frequency content: since quasibound states have $\omega\sim \mu$, it is conceivable
that high-frequency initial data just dissipates away.
Our results show that this does not happen. Instead, again, the system evolves towards a frequency content $\omega \sim \mu$ and a spatial distribution
described well by a nonrelativistic quasibound state. It seems that these quasibound states are an attractor in the phase space.

We should also add that quasibound states decay exponentially in space, far away from the binary. At large distances, the dominant behavior is controlled by power-law tails~\cite{Hod:1998ra,Koyama:2001ee,Koyama:2001qw,Witek:2012tr}. Asymptotically, our results are consistent with a late-time decay $\phi_{lm}\sim t^{-p}\sin(\mu t)$ where the exponent $p=l+3/2$ at intermediate times, and $p=5/6$ at very late times. Our results are consistent with analytical predictions~\cite{Hod:1998ra,Koyama:2001ee,Koyama:2001qw,Witek:2012tr}.

Finally, we note that if the quasibound states described above did not exist, the evolution of initial data would lead, on relatively short timescales, to the total depletion of scalar field everywhere. It would quickly be absorbed by the BHs or dispersed to infinity.

%%%%%%%%%%%%%%%%%%%%%%%%%%%%%%%%%%%%%%%%%%%%%%%%%%%%%%%%%%%%%%%%%%%%%%%%%%%%%
\section{Conclusions}
\label{sec:conclusions}
%%%%%%%%%%%%%%%%%%%%%%%%%%%%%%%%%%%%%%%%%%%%%%%%%%%%%%%%%%%%%%%%%%%%%%%%%%%%%
Light scalar fields are interesting solutions to some of the most pressing problems in physics.
One example is the dark matter problem. It is tempting to introduce fields with a scale (the Compton wavelength)
similar to the size of galactic cores. One would thus be dealing with fields of mass $10^{-21}\,{\rm eV}$ or similar,
in what are known as fuzzy dark matter models~\cite{Robles:2012uy,Hui:2016ltb}. Understanding the physics and evolution
of compact binaries in such environments is crucial to modeling their evolution and to searching for such fields~\cite{Annulli:2020ilw,Annulli:2020lyc}.

Our results indicate that, in the presence of a background scalar, the scalar field dynamics close to a BHB parallels very closely that of an electron in a 
one-electron heteronuclear diatomic molecule. 
We do note that there are very important differences between realistic BHBs and molecules. In
particular, a BHB is a dissipative system. Our numerical results for the time evolution of initial data show that
nonrelativistic bound states turn into quasibound states, via absorption at the horizon.

A BHB is dissipative in another way, not included (for simplicity) in our simulations: the system loses energy through
gravitational wave emission. We focused on timescales much shorter than the typical scale for BH
coalescence. Naturally there are systems for which this assumption is not justified.
The typical timescale until coalescence is $t_{c}\sim D^{4}/M^{3}$~\cite{Peters:1964zz} (for simplicity, we assume a circular orbit and an equal mass binary). The timescale is $t_{c}\sim 10^{7}M$ for $D=60M$, and $t_{c}\sim 10^{4}M$ for $D=10M$. Our numerical simulations clearly show that the quasibound state lifetime is at least $\mathcal{O}(10^{4})M$.
Thus, BHB evolution via gravitational-wave emission is indeed relevant for the evolution of these states, especially at small separations, and left for future work.

We have not dealt with eccentricity, nor did we consider unequal-mass binaries, although it is straightforward to apply our formalism and methods to these situations.
Unequal-mass binary evolution might lead, due asymmetric accretion and drag, to substantial center-of-mass velocities making it especially interesting to study~\cite{Cardoso:2020lxx}.

In the context of gravitational-wave imprints, dynamical friction caused by such fields and its impact on the gravitational-wave phase was recently described~\cite{Annulli:2020ilw,Annulli:2020lyc}.
However, such description does not include possible quasibound-state formation.
The clouds have a size $1/(M\mu^2)$ and according to general considerations and specific calculations~\cite{Feynman:1939zza} they should contribute an extra attracting force
which scales linearly with the scalar density. Since this is a conservative effect, its only consequence is a slight renormalization of the binary mass, and we do not expect changes to the 
dephasing introduced by dissipative effects~\cite{Annulli:2020ilw,Annulli:2020lyc} (in particular, a dephasing appearing at post-Newtonian order ``$-6$'' with respect to the leading vacuum general relativity prediction).

Still in the context of ultralight dark matter, consider a BHB evolving within a ``cloud'' of coherently oscillating scalars. This cloud could have primordial origin---and be a component of ultralight dark matter, or could simply arise as a consequence of superradiant instabilities, and be localized around a supermassive, spinning BH. 
Now, when this cloud is much larger than any scale in a BHB system, the corresponding boundary conditions are different. It is possible that molecular-like states arise, but their study requires understanding
the time evolution of scalar fields with coherent oscillating boundary conditions~\cite{Clough:2019jpm}.

% We note that the
The tidal disruption of scalar clouds by orbiting companions was recently discussed~\cite{Cardoso:2020hca}.
Our results raise the interesting possibility that the final state of such a disruption can be a gravitational molecule.

We note that the response of BHBs to external fluctuations has been studied recently. In particular, the response to high-frequency and low-frequency scalars was studied in toy models~\cite{Assumpcao:2018bka,Nakashi:2019mvs,Nakashi:2019tbz}. A realistic BHB configuration revealed already universal ringdown for binaries, and hints of superradiance~\cite{Bernard:2019nkv,Wong:2019kru,Brito:2015oca}. Together with the results we discussed here, these studies show that compact binaries are a fertile ground for new phenomenology.

Our results can also be put into the context of stationary solutions of the
field equations. One can show that minimally coupling gravity to a real scalar
field cannot give rise to stationary BH geometries with a nontrivial scalar
field---as this would give rise to a time-dependent stress-energy tensor, and
hence the emission of gravitational waves---or absorption of the scalar field at
the BH horizon. However, these results can be circumvented for single-BH
spacetimes if the field is complex (hence giving rise to a time-independent
stress-energy tensor) and the BH is spinning (where superradiance prevents
absorption by the horizon)~\cite{Herdeiro:2014goa,Brito:2015oca}. Thus,
stationary BH solutions surrounded by scalar fields are
possible~\cite{Herdeiro:2014goa}. Given the very nature of BH binaries, and the
fact that they are bound by gravity and hence evolve via gravitational-wave
emission, the existence of stationary solutions is \textit{a priori} not
expected. What we have shown, is that on timescales short compared to those
caused by energy loss through gravitational radiation, quasibound states of
scalar fields are possible and form naturally.

%%%%%%%%%%%%%%%%%%%%%%%%%%%%%%%%%%%%%%%%%%%%%%%%%%%%%%%%%%%%%%%%%%%%%%%%%%%%%
\begin{acknowledgments}

V.C.\ acknowledges financial support provided under the European Union's H2020 ERC 
Consolidator Grant ``Matter and strong-field gravity: New frontiers in Einstein's 
theory'' grant agreement no. MaGRaTh--646597.
M.Z.\ acknowledges financial support provided by FCT/Portugal through the IF
programme, grant IF/00729/2015.
T.I. acknowledges financial support provided under the European Union's H2020 ERC, Starting
Grant agreement no.~DarkGRA--757480.
This project has received funding from the European Union's Horizon 2020 research and innovation programme under the Marie Sklodowska-Curie grant agreement No 690904.
We acknowledge financial support provided by FCT/Portugal through grant PTDC/MAT-APL/30043/2017 and
through the bilateral agreement FCT-PHC (Pessoa programme) 2020-2021.
The authors would like to acknowledge networking support by the GWverse COST Action 
CA16104, ``Black holes, gravitational waves and fundamental physics.''
Computations were performed on the ``Baltasar Sete-Sois'' cluster at IST and XC40 at YITP in Kyoto University.
This work was granted access to the HPC resources of MesoPSL financed
by the Region Ile de France and the project Equip@Meso (reference ANR-10-EQPX-29-01) of the programme Investissements d'Avenir supervised
by the Agence Nationale pour la Recherche.

\end{acknowledgments}

\appendix

%%%%%%%%%%%%%%%%%%%%%%%%%%%%%%%%%%%%%%%%%%%%%%%%%%%%%%%%%%%%%%%%%%%%%%%%%%%%%
\section{Mode excitation by a binary spacetime}\label{sec:Excited mode by binary spacetime}
%%%%%%%%%%%%%%%%%%%%%%%%%%%%%%%%%%%%%%%%%%%%%%%%%%%%%%%%%%%%%%%%%%%%%%%%%%%%%
To understand which modes are excited by the BHB let us again consider
the metric in Eq.~\eqref{eq:metric}. Expanding the Newtonian potential [Eq.~(\ref{eq:newton})],
we obtain
\begin{align}
  \Phi 
  % -\frac{M_{1}}{|\vec{r}-\vec{r}_{1}(t)|}-\frac{M_{2}}{|\vec{r}-\vec{r}_{2}(t)|}\\
&\simeq-\frac{M_{1}+M_{2}}{r}+\frac{M_{1}r_{1}(t)^{2}+M_{2}r_{2}(t)^{2}}{2r^{3}}\nonumber\\
&-\frac{M_{1}(\vec{r}\cdot\vec{r}_{1}(t))^{2}+M_{2}(\vec{r}\cdot\vec{r}_{2}(t))^{2}}{2r^{5}}\,.
\end{align}
The Klein-Gordon equation on this spacetime can be written as
\begin{eqnarray}
\left(-\partial_{t}^{2}+\nabla^{2}
-\mu^{2}\right)\phi
=
-4\partial_{t}\Phi\partial_{t}\phi-4\Phi\nabla^{2}\phi+2\mu^{2}\Phi\phi\,.\nonumber
\end{eqnarray}
%
%Since left hand side is massive Klein-Golden operator, 
The formal solution of this equation is
\begin{eqnarray}
\phi-\phi^{\rm hom}&=&\int dt^{\prime}d^{3}\vec{x}^{\prime}G(t-t^{\prime},\vec{x}-\vec{x}^{\prime})\nonumber\\
&\times&\left(-4\partial_{t'}\Phi\partial_{t'}\phi-4\Phi\nabla'^{2}\phi+2\mu^{2}\Phi\phi\right)(t',\vec{x}')\nonumber
\label{Born approximated eq}
\end{eqnarray}
where $\phi^{\rm hom}$ is the homogeneous solution, fixed by the initial data.
$G(t-t^{\prime},\vec{x}-\vec{x}^{\prime})$ is the Green's function,
\begin{eqnarray}
G(t,\vec{x})&=&\int \frac{d^{3}\vec{k}}{(2\pi)^{3}2\omega_{k}}e^{-i\omega_{k}t+i\vec{k}\cdot\vec{x}}i \, \theta(t)+{\rm c.c.}\,,
\end{eqnarray}
where $\omega_{k}=\sqrt{\mu^{2}+\vec{k}^{2}}$,
and ${\rm c.c.}$ is a complex conjugate.
For weak couplings, $\phi \sim \phi^{\rm hom}$,
\begin{align}
&\phi-\phi^{\rm hom}=\int dt^{\prime}d^{3}\vec{x}^{\prime}G(t-t^{\prime},\vec{x}-\vec{x}^{\prime}) \\
&\times \big(-4\partial_{t'}\Phi\partial_{t'}\phi^{\rm hom}-4\Phi\nabla'^{2}\phi^{\rm hom} +2\mu^{2}\Phi\phi^{\rm hom}\big)(t',\vec{x}')\nonumber\,.
\end{align}

Let us now expand the right-hand side of this equation in spherical harmonics,
\begin{align}
\phi^{\rm hom}&=\sum_{lm}\phi^{{\rm hom}}_{lm}(t,r)Y^m_{l}(\theta,\varphi)\,,\\
\Phi&=\sum_{lm}\Phi_{lm}(t,r)Y^m_{l}(\theta,\phi) =\Phi_{00}(t,r)Y^0_{0}(\theta,\phi) \nonumber \\
&\quad+\sum_{l=2,4,6,\cdots}\sum_{m=\pm 2}\Phi_{lm}(t,r)Y^m_{l}(\theta,\phi)\,,
\label{eq:Phi expansion}
\end{align}
where
\begin{align*}
\Phi_{00}&=\frac{M_{1}+M_{2}}{r}+\frac{M_{1}R_{1}^{2}+M_{2}R^{2}_{2}}{2r^{3}}-\frac{3}{4}\frac{M_{1}R_{1}^{2}+M_{2}R_{2}^{2}}{r^{3}} \\
\Phi_{l,\pm2}&=-\frac{3}{2}\frac{M_{1}R_{1}^{2}+M_{2}R_{2}^{2}}{r^{3}}c_{l}e^{\mp2i\Omega t}
\end{align*}
and
$c_{2}=\frac{1}{2}\sqrt{\frac{5\pi}{6}}$, $c_{4}=\frac{1}{2}\sqrt{\frac{\pi}{10}}$, $c_{6}=\frac{1}{4}\sqrt{\frac{13\pi}{105}}$, $\ldots$.
The Green function can also be expanded in spherical harmonics through the plane wave expansion
\[
e^{i\vec{k}\cdot\vec{x}}=4\pi\sum_{l=0}^{\infty}\sum_{m=-l}^{l}i^{l}j_{l}(kr)Y^m_{l}(\hat{\vec{k}})Y^m_{l}{}^{\ast}(\hat{\vec{x}}) \,,
\]
where the $j_{l}$ are the spherical Bessel functions and the hat $\hat{}$ denotes a unit vector.
Thus,
\begin{widetext}
\begin{align*}
\phi(t,\vec{x})-\phi^{\rm hom}(t,\vec{x})
&=\int dt'd^{3}\vec{x}'\frac{d^{3}\vec{k}}{(2\pi)^{3}2\omega_{k}}\biggl\{
e^{-i\omega_{k}(t-t')}i\,\theta(t-t') \, e^{i\vec{k}\cdot\vec{x}}\\
&\sum_{l,m}\sum_{l',m'}\sum_{l'',m''}
4\pi (-i)^{l''}j_{l''}(kr')Y^{m''}_{l''}{}^{\ast}(\hat{\vec{k}})Y^{m''}_{l''}(\theta',\varphi')Y^m_{l}(\theta',\varphi')Y^{m'}_{l'}(\theta',\varphi')\mathcal{A}_{lml'm'}(t',r')
+{\rm c.c.}
\biggr\}\,,
\end{align*}
% \end{widetext}
where
% \begin{widetext}
%
\begin{align*}
\mathcal{A}_{lml'm'}(t,r)&=
-4\partial_t\Phi_{lm}(t,r)\partial_{t}{\phi}^{\rm hom}_{l'm'}(t,r)
-4\Phi_{lm}(t,r)\left(
\partial_{rr}\phi^{\rm hom}_{l'm'}(t,r)
+\frac{2}{r}\partial_r\phi^{\rm hom}_{l'm'}(t,r)
-\frac{l'(l'+1)}{r^{2}}\phi^{\rm hom}_{l'm'}(t,r)
\right)\\
&\quad+2\mu^{2}\Phi_{lm}(t,r) \phi^{\rm hom}_{l'm'}(t,r)\,.
\end{align*}
%
% \end{widetext}

Let us consider the modes of the scattered field,
$\left(
\phi(t,\vec{x})-\phi^{\rm hom}(t,\vec{x})
\right)_{lm}$.
Performing the integration in the angular directions of $\vec k$ we obtain
% \begin{widetext}
\begin{align}
\left(
\phi(t,\vec{x})-\phi^{\rm hom}(t,\vec{x})
\right)_{lm}&=
\int \frac{dt' dr' r'^{2} dk \, k^{2}}{\pi\omega_{k}} 
e^{-i\omega_{k}(t-t')} i \, \theta(t-t') \, j_{l}(kr) \, j_{l}(kr') \notag\\
& \qquad
\sum_{l'm'}\sum_{l''m''}
\left\{
\Lambda_{l,l',l''}^{m,m',m''}
\mathcal{A}_{l''m''l'm'}(t',r')
+ (-)^{m} \Lambda_{l,l',l''}^{-m,m',m''}
\mathcal{A}_{l''m''l'm'}(t',r')
\right\}\,,
\end{align}
where
\begin{align*}
\Lambda_{l,l',l''}^{m, m', m''}\equiv\int d\Omega \, Y^{m}_{l}(\theta,\varphi) Y_{l'}^{m'}(\theta,\varphi) Y_{l''}^{m''}(\theta,\varphi) \,.
\end{align*}
This integral is nonzero only when $m' + m = -m''$.
Since, from Eq.~\eqref{eq:Phi expansion}, $\mathcal{A}_{l''m''l'm'}(t,r)$ is nonzero only for $m'' = 0,\,\pm 2$
%for $(l'',m'')=(0,0),\, (2,\pm 2), \, (4,\pm 2), \, \ldots$,
we see that
$\left(
\phi(t,\vec{x})-\phi^{\rm hom}(t,\vec{x})
\right)_{lm}$ is nontrivial only when
$m=\pm m'$, $m=m'\pm 2$ or $m=-m'\pm 2$, for nontrivial values of $\phi^{\rm hom}_{l'm'}$.
\end{widetext}

%%%%%%%%%%%%%%%%%%%%%%%%%%%%%%%%%%%%%%%%%%%%%%%%%%%%%%%%%%%%%%%%%%%%%%%%%%%%%
\section{Convergence test}\label{sec:Convergence test}
%%%%%%%%%%%%%%%%%%%%%%%%%%%%%%%%%%%%%%%%%%%%%%%%%%%%%%%%%%%%%%%%%%%%%%%%%%%%%

% Here, we illustrate numerical convergence of our code.
As mentioned in Sec.~\ref{sec:implementation}, for our numerical implementation
we approximate spatial derivatives with fourth-order-accurate finite difference
stencils and integrate in time with a fourth-order Runge-Kutta scheme.
Communication between refinement levels is done by Carpet with second- and
fifth-order accuracy in time and space, respectively.
% Thereby, we expect the numerical error is proportional to square of grid resolution.

\begin{figure}[tbh]
\includegraphics[clip, width=8.0cm]{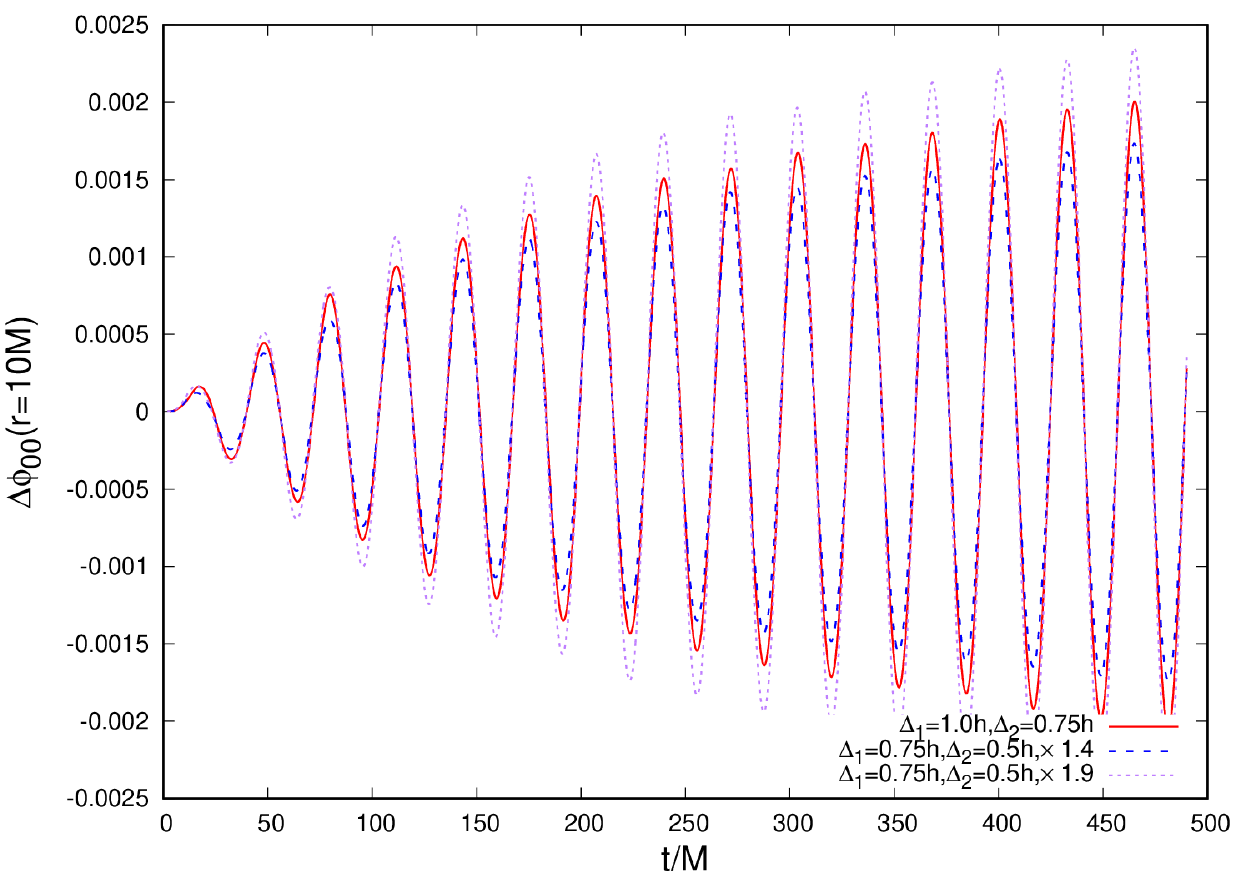}
\caption{Convergence analysis of the $l=0$, $m=0$ multipole of $\phi$, extracted at $r=10M$, for simulation \texttt{nonspin2}. The blue line shows the expected result for second-order convergence ($Q_{2}=1.4$), while the purple line is the expected result for third-order convergence ($Q_{3}=1.9$).
\label{convergence}}
\end{figure}

To assess the convergence properties of our results we performed three
simulations for configuration \texttt{nonspin2}, with resolutions (on the
coarsest refinement level) $\Delta_{c}=1.0M$, $\Delta_{m}=0.75M$, and $\Delta_{h}=0.5M$. We
define the usual convergence factor
\begin{eqnarray}
Q_{n}=\frac{f_{\Delta_{c}}-f_{\Delta_{m}}}{f_{\Delta_{m}}-f_{\Delta_{h}}}=\frac{\Delta_{c}^{n}-\Delta_{m}^{n}}{\Delta_{m}^{n}-\Delta_{h}^{n}} \,,
\end{eqnarray}
where $n$ is the expected convergence order, and depict the corresponding
results for the $l=0$, $m=0$ multipole of the evolved function $\phi$ in
Fig.~\ref{convergence}. The results are compatible with a convergence order
between orders 2 and 3, which is consistent with the numerical scheme used.

\bibliographystyle{apsrev4-1}
\bibliography{References}

\end{document}